\documentclass[12pt,a4paper]{article}
\usepackage[utf8]{inputenc}
\usepackage[T1]{fontenc}
\usepackage[british,english]{babel}
\usepackage{graphicx}
\usepackage{epstopdf}
\usepackage[margin=2.5cm]{geometry}
\usepackage[labelfont=bf]{caption}
\usepackage[round]{natbib}
\usepackage{booktabs}
\usepackage{tabularx}
\usepackage{setspace}
\usepackage{pdflscape}
\usepackage{amsmath,amssymb}
\usepackage{hyperref}
\usepackage{verbatim}
\usepackage{fancyhdr}
\usepackage{beramono}
\usepackage{listings}
\usepackage[usenames,dvipsnames]{xcolor}
\usepackage[displaymath, mathlines,running]{lineno}
\usepackage[colorinlistoftodos]{todonotes}
\usepackage{caption}
\usepackage{subcaption}
\usepackage{adjustbox}

\usepackage{authblk}
\usepackage{fancyhdr}

\hypersetup{
    colorlinks,
    linkcolor={red!50!black},
    citecolor={blue!50!black},
    urlcolor={blue!80!black}
}

\title{Increased Electrification of Heating and Weather Risk in the Nordic Power
  System}
\author[1,2]{Ian M. Trotter\thanks{Corresponding author. E-mail: \href{mailto:ian.trotter@ufv.br}{\texttt{ian.trotter@ufv.br}}}}
\author[2]{Torjus F. Bolkesjø}
\author[2]{Eirik O. Jåstad}
\author[2]{Jon Gustav Kirkerud}
\affil[1]{\normalsize Institute of Public Policy and Sustainable Development/Department of
  Agricultural Economics, Universidade Federal de Viçosa}
\affil[2]{Faculty of Environmental Sciences and Natural Resource Management, Norwegian University of Life Sciences}

\date{\textbf{Working Paper} -- December 2021}
    
\begin{document}
\maketitle

\begin{abstract}
  \noindent
  Weather is one of the main drivers of both the power demand and supply,
  especially in the Nordic region which is characterized by high heating needs
  and a high share of renewable energy.
  Furthermore, ambitious decarbonization plans may cause power to replace
  fossil fuels for heating in the Nordic region, at the same time as large wind
  power expansions are expected, resulting in even greater exposure to weather
  risk.
  In this study, we quantify the increase in weather risk resulting from
  replacing fossil fuels with power for heating in the Nordic region, at the
  same time as variable renewable generation expands.
  First, we calibrate statistical weather-driven power consumption models for
  each of the countries Norway, Sweden, Denmark, and Finland.
  Then, we modify the weather sensitivity of the models to simulate different
  levels of heating electrification, and use 300 simulated weather years to
  investigate how differing weather conditions impact power consumption at each
  electrification level.
  The results show that full replacement of fossil fuels by power for heating
  in 2040 leads to an increase in annual consumption of 155 TWh (30\%) compared
  to a business-as-usual scenario during an average weather year, but a 178 TWh
  (34\%) increase during a one-in-twenty weather year.
  However, the increase in the peak consumption is greater: around 50\% for a
  normal weather year, and 70\% for a one-in-twenty weather year.
  Furthermore, wind and solar generation contribute little during the
  consumption peaks.
  The increased weather sensitivity caused by heating electrification causes
  greater total load, but also causes a significant increase in inter-annual,
  seasonal, and intra-seasonal variations.
  We conclude that heating electrification must be accompanied by an increase in
  power system flexibility to ensure a stable and secure power supply.
\end{abstract}

\section{Introduction}
Weather is one of the main drivers of both demand and supply in the power
sector, and its impact is increasing \citep{mideksa_impact_2010,
  staffell_increasing_2018}.
On the demand side, weather conditions mainly affect power consumption related
to heating and cooling \citep{dryar_effect_1944, quayle_heating_1980,
  hor_analyzing_2005, trotter_climate_2016, rodriguez_climate_2019}.
On the supply side, weather directly determines wind \citep{foley_current_2012}
and solar power generation \citep{shi_forecasting_2012,
  pfenninger_long-term_2016, sanjari_probabilistic_2017}, as well as the
conditions for hydropower production \citep{kaunda_hydropower_2012,
  birkedal_determinants_2016}.
Therefore, weather conditions are of fundamental importance in the power
sector, and represent a significant source of variation and uncertainty.

\begin{figure}
  \centering
  \includegraphics[width=\textwidth]{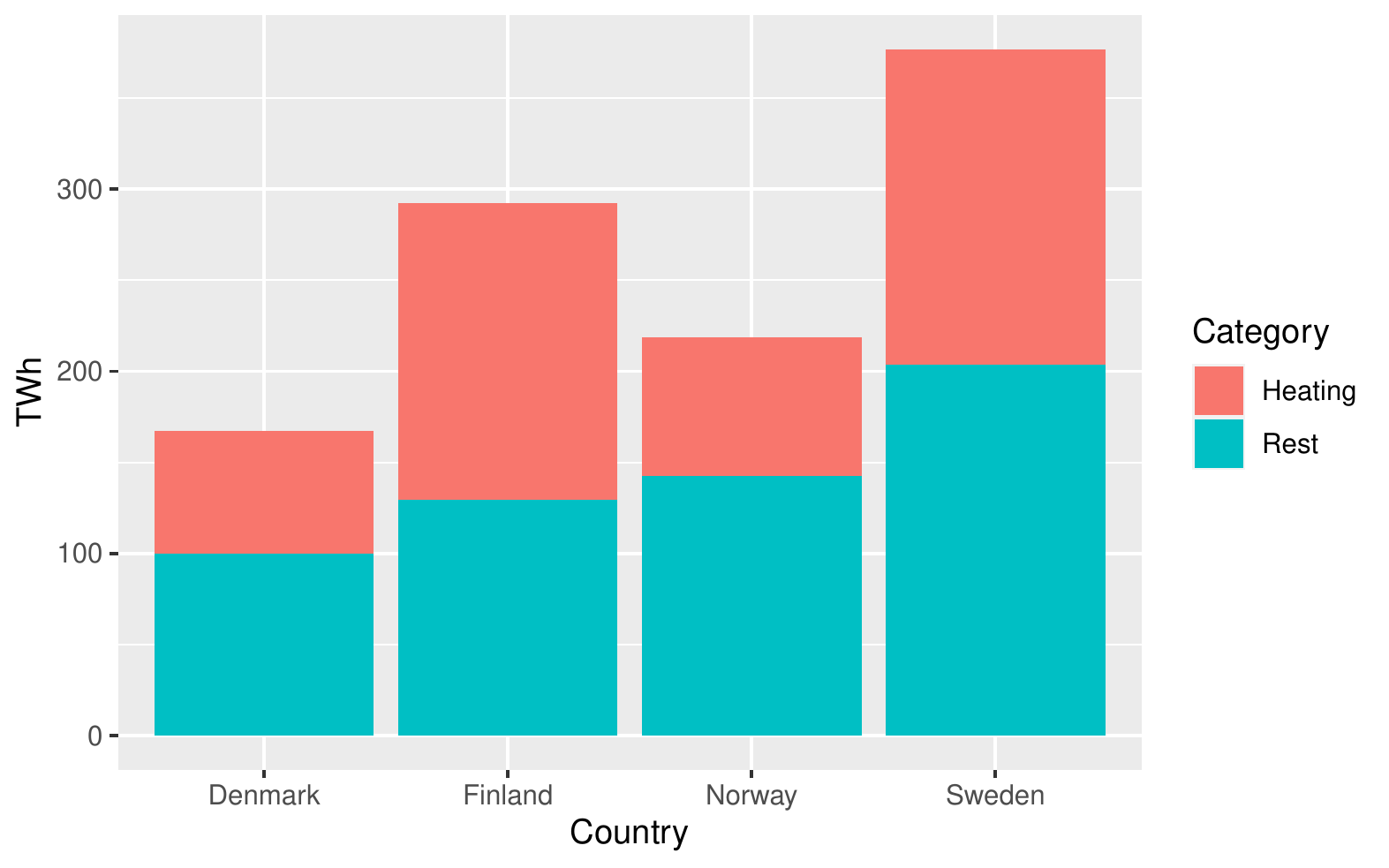}
  \caption{Heating energy demand in total final energy demand for the Nordic
    countries in 2012. Based on data by \cite{fleiter_mapping_2016}.}
  \label{fig:heatingshare}
\end{figure}

\begin{figure}
  \centering
  \includegraphics[width=\textwidth]{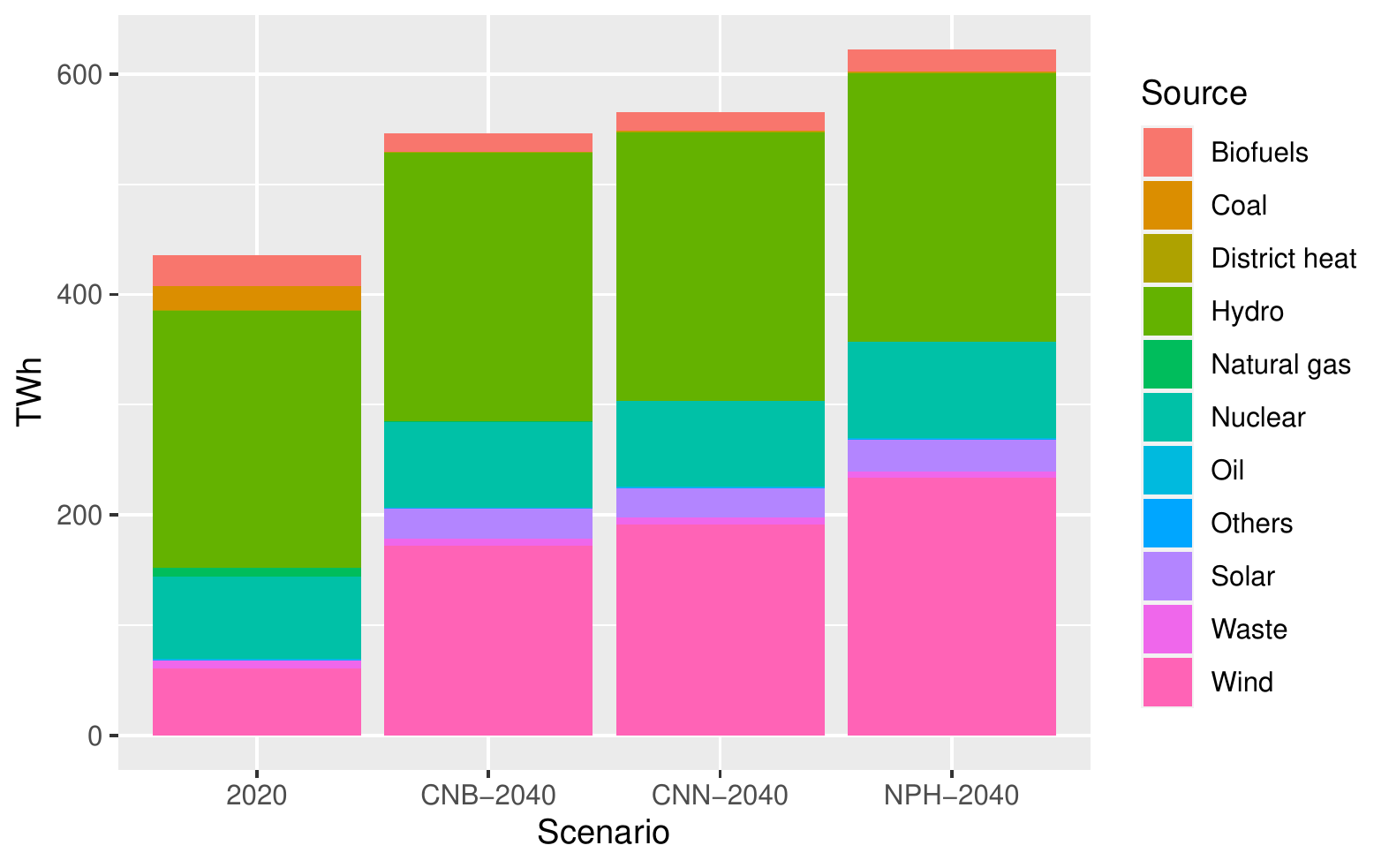}
  \caption{Nordic installed power generation capacity for 2020, and projections
    2040 for the Carbon Neutral Behaviour (CNB), Carbon Neutral Nordic (CNN),
    and Nordic Power House (NPH) scenarios of \cite{wrake_nordic_2021}.}
  \label{fig:electricitygenerationsources}
\end{figure}

At the same time, two ongoing developments may further increase the Nordic power
system's exposure to weather risk.
Firstly, the transition away from fossil fuels may result in an increase in
electric heating.
Due to the cold climate in the Nordic countries (Norway, Sweden Denmark and
Finland), these countries already consume a large amount of energy for heating
purposes -- approximately 480 TWh in 2012, representing almost half the final
energy demand \citep{fleiter_mapping_2016}, illustrated in Figure
\ref{fig:heatingshare}.
In the four countries combined, 17\% of the heating energy in 2012 was provided
by electricity and 28\% was provided directly by fossil fuels (such as coal,
fuel oil and natural gas), but also 85\% of the installed district heating
capacity, which supplies around 21\% of heating energy, relies directly or
indirectly on fossil fuels \citep{fleiter_mapping_2016}.
\begin{figure}
  \centering
  \includegraphics[width=\textwidth]{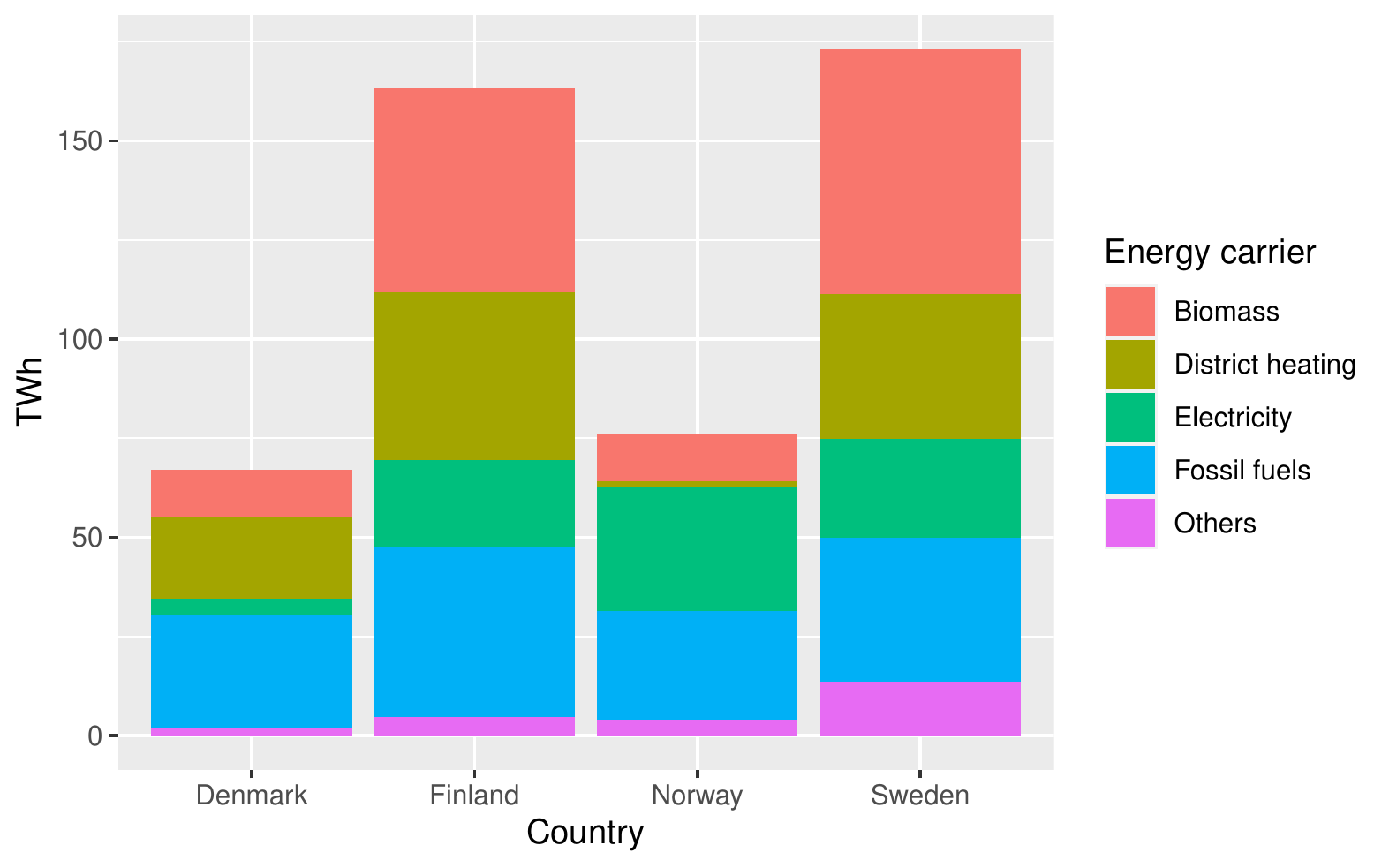}
  \caption{Energy carriers of final energy demand used for heating in the Nordic
    countries in 2012. Based on data by \cite{fleiter_mapping_2016}.}
  \label{fig:heatingenergycarriers}
\end{figure}
Therefore, there is a large potential for increased electric heating in the
region, especially in the context of a decarbonisation of the energy sector.
As a result of increased reliance on electricity as an energy carrier for
heating, the power consumption will become increasingly sensitive to weather
conditions.
Secondly, the share of intermittent renewable generation capacity, such as wind
and solar power, is increasing as part of a transition to renewable energy
sources.
\cite{iea_nordic_2016} expect wind power generation in the Nordics to increase
around five-fold from 2013 to 2050, and \cite{wrake_nordic_2021} project an
increase in solar and wind power generation of around three- or four-fold from
2020 to 2040, as illustrated in Figure \ref{fig:electricitygenerationsources}.
Since wind and solar power generation are not \emph{dispatchable} -- that is,
the operators cannot generally choose whether or not to generate electricity at
a given moment -- the direct and immediate impacts of weather conditions on
power generation also tend to increase as the share of intermittent renewables
increases.
As a result of these two developments, the Nordic power system may experience an
increased exposure to weather risk in the future, which is a cause of concern
for system operators, power market participants, investors, regulators, and
policymakers.
In addition, the Nordic countries were amongst the first countries in Europe to
deregulate their power sectors, and risk management in a competitive,
deregulated market can have consequences for market stability and resilience.
In order to ensure a stable and secure power supply, it is therefore important
to understand how these developments may affect the weather risk faced by the
Nordic power system.

In this study, we therefore investigate the impact of heating electrification on
the weather risks for the Nordic power system -- Norway, Sweden, Denmark and
Finland -- and analyse its implications, in light of the ongoing energy
transition.
First, we calibrate statistical power consumption models for these countries
that capture the historical relationships between weather and power consumption.
Second, we modify the temperature sensitivity of the consumption models to
represent different levels of heating electrification.
In the third step, we use these models to generate consumption projections for
the year 2040, using 300 simulated weather years, and also pair the consumption
projections with projections for intermittent power generation.
This allows us to analyse the impact of heating electrification on consumption
under many possible weather conditions.
Furthermore, our analysis will also show how power consumption will interact
with intermittent power generation, considering the large capacity expansion
expected until year 2040.
Taken together, this will provide a deeper understanding of how the
electrification of heating will affect the weather risk of the Nordic power
system, within the context of the ongoing energy transition.

Heating electrification has received increasing research attention lately, as
many regions are looking to replace fossil energy with renewable power in order
to lower greenhouse gas emissions in response to climatic change.
Not only will heating electrification normally lead to higher power consumption
\citep{watson_decarbonising_2019}, but it will also alter the profile of the
power consumption \citep{veldman_impact_2011, staffell_increasing_2018}.
Larger amounts of electric heating would result in power consumption becoming
more weather sensitive, which again would cause greater variability and
uncertainty in the power consumption \citep{wilson_historical_2013}, and may
increase the frequency of unservicable deficits
\citep{quiggin_implications_2016}.
This would translate into increased variability in system costs, especially in
systems with high VRE share \citep{heinen_electrification_2017}, and greatly
increase the need for power system flexibility \citep{staffell_increasing_2018,
  thomasen_decarbonisation_2021}.
Despite these challenges, electrification of heating coupled with a large
expansion in VRE generation is considered a viable -- and even promising --
strategy for reducing emissions in several regions
\citep{kirkerud_power--heat_2017, sheikh_decarbonizing_2019,
  ruhnau_heating_2020, chen_impact_2021, sakamoto_demand-side_2021}, although it
may only be cost-efficient if emissions are relatively costly
\citep{haghi_iterative_2020}.

In the context of the literature on heating electrification, our study makes two
main contributions. Firstly, we focus on the Nordic region, which has not been
the main focus of any earlier heating electrification studies. Although the
Nordic region has been included in the study on Northern Europe by
\cite{chen_impact_2021} and partly considered by
\cite{thomasen_decarbonisation_2021}, there are compelling reasons to focus
exclusively on this region. Due to the harsh climate in this region, heating is
a basic need and the heating requirements are relatively large. In addition, the
Nordic region has ambitious emissions reductions targets, as the countries
target carbon neutrality between 2030 and 2050\footnote{The target year for
  carbon neutrality in Norway is 2030, 2035 in Finland, 2045 in Sweden, and 2050
  in Denmark.} \citep{wrake_nordic_2021}. This makes the Nordic region
particularly interesting to study in the context of heating
electrification. Secondly, none of the earlier heating electrification studies
focus explicitly on weather risk, even though most agree that it is of great
interest. Some studies incorporate some degree of weather variability by
considering several historical weather years -- such as
\cite{quiggin_implications_2016}, \cite{heinen_electrification_2017},
\cite{staffell_increasing_2018} and \cite{watson_decarbonising_2019} -- but we
devote greater attention to this aspect than previous studies on heating
electrification by estimating outcome densities using 300 simulated weather
years.

Weather risk in power systems has been extensively studied previously, outside
the context of heating electrification. The professional community was early to
incorporate weather into load forecasting \citep{dryar_effect_1944,
  heinemann_relationship_1966}. The meteorological community was also early to
study the link between weather and power demand \citep{thom_rational_1954,
  quayle_heating_1980}, and the econometric community quickly followed
\citep{fisher_study_1962, halvorsen_residential_1975}.
Lately, the focus has gradually shifted from point forecasts to probabilistic
forecasts, which in some sense represent weather risk
\citep{veall_boostrapping_1987, mcsharry_probabilistic_2005,
  hyndman_density_2010, sideratos_probabilistic_2012, tastu_probabilistic_2014,
  trotter_climate_2016, wang_deep_2017}.
As climatic change gained importance on the research agenda, the forecasting
horizon also increased from hours, days or weeks to decades
\citep{parkpoom_climate_2004, hyndman_density_2010, trotter_climate_2016,
  fan_impacts_2019, silva_climate_2020}.
A growing number of studies have also been concerned with modelling and
forecasting wind power generation \citep{foley_current_2012,
  tobin_assessing_2015, kiviluoma_variability_2016, garrido-perez_impact_2020}
and solar power generation \citep{shi_forecasting_2012, wild_projections_2015,
  pfenninger_long-term_2016, sanjari_probabilistic_2017,
  castillejo-cuberos_understanding_2020}. While earlier studies often
considered each of these elements in isolation, many recent studies analyse
various elements in combination, such as wind and solar
\citep{heide_seasonal_2010, troccoli_large-scale_2010, widen_correlations_2011,
  jerez_spatio-temporal_2013, bett_climatological_2016,
  solomon_investigating_2016, miglietta_local_2017}, solar and hydro
\citep{siala_solar_2021}, wind and load \citep{sinden_characteristics_2007,
  leahy_wind_2012, baringo_correlated_2013, coughlin_multi-scale_2014,
  bell_wind_2015, thornton_relationship_2017}, or even wind, solar and hydro
\citep{canales_assessing_2020}. Some studies incorporate additional elements,
including the impacts of weather on thermoelectric power plants
\citep{tobin_vulnerabilities_2018}, prices \citep{suomalainen_correlation_2015},
and tidal power generation
\citep{coker_measuring_2013}. \cite{engeland_space-time_2017} and
\cite{widen_variability_2015} present more comprehensive reviews of the
literature on the variability of renewable power generation.
In a particularly interesting pair of studies,
\cite{van_der_wiel_meteorological_2019} use a large number of weather
simulations to investigate the risk of extreme shortfalls between renewable
power production and demand in Europe, and \cite{van_der_wiel_influence_2019}
further establish that the risk increases during blocked circulation patterns,
such as ``Scandinavian blocking'' and ``North Atlantic Oscillation
negative''. Interestingly, the authors show that changes due to climate change
are substantially smaller than interannual weather variability.
Further, focusing on the meteorological variables,
\cite{ramsebner_estimating_2021} also explore the correlations between renewable
generation and proxies for heating/cooling needs in Europe.
These studies all concern the impact of weather on the power system. Although
some of the studies present long-term projections (such as
\cite{hyndman_density_2010}, \cite{trotter_climate_2016} and
\cite{rodriguez_climate_2019}) and even specifically investigate weather risk
(such as \cite{mcsharry_probabilistic_2005} and
\cite{van_der_wiel_meteorological_2019}), the studies in this line of research
have not yet addressed the question of heating electrification.

Compared to the existing literature on weather risk in power systems, which
generally implicitly assumes that weather sensitivity remains constant, our
study therefore contributes by simulating scenarios where the weather
sensitivity of the power system increases. Our study not only calibrates the
statistical models and presents projetions, but also modifies the models to
represent fundamental changes in the underlying reality -- the electrification
of heating -- and thereby creating and comparing alternate scenarios. As such,
our study contributes to a deeper understanding of how heating electrification
may affect weather risk, and of the future of the Nordic power system in
particular, which is relevant to researchers, policymakers and market
participants -- particularly in light of the ongoing energy transition.

This study is divided into four sections, including this introduction. The
following section details the methodology of our investigation, whereas the
third section presents, synthesizes and discusses the results of our
experiment. The fourth and final section summarizes the main conclusions of the
investigation, and offers suggestions for future investigations.

\section{Methodology}
The objective of this study is to analyse the weather risks of the Nordic power
system, under conditions of increased heating electrification and increased
variable renewable power generation. Our strategy to achieve this consists of
three main steps:

First we calibrate power consumption models for each of the Nordic countries --
Norway, Sweden, Denmark and Finland -- using historical consumption and weather
data. Throughout, we work at an hourly resolution so that we capture intra-day
variations in both consumption and variable renewable generation.

Secondly, we modify the calibrated power consumption models in order to simulate
increased levels of heating electrification in the Nordic countries for the year
2040.

Thirdly, we use the consumption models together with a large amount of simulated
weather scenarios to generate a large number of possible joint paths for power
consumption and variable renewable generation for 2040. This will show how
differing weather conditions impact the power consumption and variable renewable
power generation at different levels of heating electrification, and allow us to
estimate the density functions of key figures such as total annual power
consumption, annual peak power consumption, and annual peak residual demand,
which is the remainder when we subtract wind and solar generation from
consumption. Analysing the load duration curves of the different electrification
scenarios will also provide further insight into how heating electrification
impacts weather risk.

We now explain each step of our methodology in greater detail.

\subsection{Consumption Model Calibration}
\label{sec:consmodels}
A separate consumption model will be calibrated for each of the four countries
-- Norway, Sweden, Denmark and Finland -- at hourly resolution. The consumption
models will relate hourly power consumption to the temperature at a set of $n=5$
weather stations in each country through heating degree hours (HDH) and cooling
degree hours (CDH), with cut-off temperatures of 17$^\circ$C and 22$^\circ$C,
respectively. We further include two important socio-economic indicators in the
models: gross domestic product (GDP) and population (POP). In order to capture
seasonalities at different timescales, we include dummy variables for each hour
of the day (HR), each month of the year (MTH), each day of the week (WD), and an
indicator for holidays (HOL), as well as a trend variable (T). We base the model
on the natural logarithm of consumption, income and population, and calibrate
the model using ordinary least squares. As such, the estimated models can be
represented in the following functional form:
\begin{linenomath}
\begin{align*}
  \ln(\text{Cons}_{t}) =& a \ln(\text{GDP}_t) + b \ln(\text{POP}_t) \\ & +
  \sum_{i=1}^n c_i \text{HDH}^i_t + \sum_{i=1}^n d_i \text{CDH}^i_t \\ & +
  \sum_{i=1}^{24} e_i \text{HR}^i_t + \sum_{i=1}^{12} f_i \text{MTH}^i_t +
  \sum_{i=1}^7 g_i \text{WD}_{i,t} + h \text{T}_t + j \text{HOL}_t.
\end{align*}
\end{linenomath}
The data is first split into training and validation samples, consisting of 75\%
and 25\% of the sample. We calibrate the models on the training samples using
ordinary least-squares regression, and measure the in-sample accuracies, as well
as the out-of-sample accuracies using the validation sample. The out-of-sample
accuracies will give an indication of how the models perform on data outside of
the training set.

To measure the model accuracies, we calculate several error indices: root mean
squared error (RMSE), mean absolute error (MAE), mean absolute percentage error
(MAPE) and symmetric mean absolute percentage error (sMAPE). We then recalibrate
the model on the full dataset, and measure the in-sample accuracy, before using
the models for creating projections. In addition, we measure the relative
importance of each group of model inputs by estimating the effect sizes with
Cohen's $f^2$ \citep{selya_practical_2012}, in order to verify that weather is
in fact an important determinant of power consumption.

\subsection{Electrification and VRE Scenario Design}
\label{sec:elecscen}
To simulate increased heating electrification, we modify the coefficients
related to the power consumption for heating purposes, specifically the
coefficients for heating degree hours, HDH.
By applying a certain percentage increase to the HDH coefficient, we increase
the weather-sensitivity of the power consumption, which is one of the main
effects of heating electrification on power consumption (see, for instance,
\cite{wilson_historical_2013, quiggin_implications_2016,
  heinen_electrification_2017, thomasen_decarbonisation_2021}).
Based on available data from 2012 \citep{fleiter_mapping_2016}, we calculate the
percentage to increase the HDH coefficients such that they represent the
replacement of a certain proportion of the fossil-based heating.
Although more recent data may be available, we design the electrification
scenarios based on numbers from 2012 because it was a fairly typical year for
the Nordic power sector, and because it is within the later part of the
calibration period of the consumption models, ensuring that the designed
scenarios are consistent with the calibrated consumption models.
This means, however, that some recent developments, such as the recent increases
in the use of biofuels for heating, are not entirely reflected in the scenarios.

In order to calculate the percentage increase for the HDH coefficients, we must
first determine how much power will be needed to replace fossil-based heating.
We distinguish between three main uses of fossil fuels in the heating sector:
direct heating of space and water, use in district heating, and for process
heat.
We assume that space/water and district heating are temperature sensitive,
whilst process heat is not.
Therefore, only replacement of fossil fuels in space/water and district heating
contributes to increasing the temperature-sensitivity of power demand.
In addition, some of the fossil fuel heating may be replaced by electric heat
pumps, leading to efficiency gains.
Assuming that fossil fuels for heating have around 90\% efficiency, and that
75\% of the temperature-sensitive fossil fuel heating is replaced by heat pumps
with a coefficient of performance (COP) of 3 \citep{wilson_historical_2013},
whereas 25\% is replaced by resistive heating, then 1 J of fossil fuel can be
replaced by 0.475 J of electricity.
Table \ref{tab:heatingelectrification} shows the amount of fossil space/water
and district heating in each of the Nordic countries in 2012
\citep{fleiter_mapping_2016}.
By multiplying the sum of these by 0.475, we find how much electricity will be
needed to replace the temperature sensitive fossil-based heating.
Comparing the replacement electricity to the direct electric heating, we find
how large a percentage increase this would imply for the temperature
sensitivity.
For instance, we assume that 5.0 TWh of electricity could replace 10.6 TWh of
temperature sensitive fossil heating (space/water and district heating) in
Norway, which would represent a 17.4\% increase in existing direct electric
space/water heating (29.0 TWh).

\begin{table}
  \centering
  \caption{Calculation of consumption model modifications. Temperature sensitive
    fossil-based heating (space/water and district heating) could be replaced by
    a combination of electric heat pumps (75\%) with a COP of 3 and electric
    resistive heating (25\%). Such a replacement implies that temperature
    sensitive power consumption would increase by a given percentage to provide
    the additional power. Based on data by \cite{fleiter_mapping_2016}.}
  \label{tab:heatingelectrification}
  \begin{adjustbox}{width=\textwidth}
  \begin{tabular}{|l|r|r|r|r|}
    \hline
    \textbf{} & \textbf{Norway} & \textbf{Sweden} &
    \textbf{Denmark} & \textbf{Finland} \\
    \hline
    Fossil Space/Water Heating    &  9.7 TWh & 11.0 TWh & 16.9 TWh & 17.9 TWh \\
    Fossil-based District Heating &  0.9 TWh &    0 TWh & 18.1 TWh & 36.4 TWh \\
    \hline
    \textbf{Total Temp. Sensitive Fossil}  & 10.6 TWh & 11.0 TWh & 35.0 TWh & 54.3 TWh \\
    Replacement Electric          &  5.0 TWh &  5.2 TWh & 16.6 TWh & 25.8 TWh \\
    \hline
    Direct Electric S/W Heating   & 29.0 TWh & 21.5 TWh &  2.7 TWh & 20.0 TWh \\
    \hline
    \textbf{Implied Temp. Sensitivity Increase}& 17.4\% &   24.3\% &  615.7\% &  129.0\% \\
    \hline
    Fossil Process Heat           & 17.6 TWh & 25.4 TWh & 11.9 TWh & 24.9 TWh \\
    \hline    
  \end{tabular}
  \end{adjustbox}
\end{table}

We assume that process heat will be replaced with electricity at the same
efficiency rate, with a flat profile over the year.

Based on these assumptions, we define three different scenarios for 2040,
characterised by how large share of fossil fuels are substituted by power in the
heating sector:
\begin{description}
  \item[Business-As-Usual (BAU)] is a baseline scenario that will serve as a
    basis for comparison. In this scenario, the temperature sensitivity of the
    original consumption models will remain unchanged, and no additional process
    heat is explicitly electrified.
  \item[Half Electrification (HALF)] assumes that half of the heating from
    fossil fuels will be replaced by electric heating in 2040. Table
    \ref{tab:electrificationscenarios} shows the increase in temperature
    sensitivity and the additional baseload applied to the power consumption
    models for this scenario.
  \item[Full Electrification (FULL)] assumes that all the heating that was based
    on fossil fuels in 2012 is electrified, again with 75\% of the temperature
    sensitive heating being replaced by heat pumps and 25\% being replaced by
    resistive heating, and the process heat being replaced with a flat
    consumption over the year. This implies, as shown in Table
    \ref{tab:electrificationscenarios}, an increase in the temperature
    sensitivity for Norway of 17.4\%, Sweden of 24.3\%, Denmark of 615.8\%, and
    Finland of 129\%.
\end{description}
These scenarios are highly stylized and deliberately simplify the possible range
of outcomes of the energy transition, as well as many technical and economic
aspects regarding the efficiency and adoption of heat pumps and resistive
heating. Nonetheless, we believe that the simplicity of the scenarios will serve
to draw clearer insights from this thought experiment, and that these scenarios
are capable of representing and illustrating the potential impacts in a broad
sense.

\begin{table}
  \centering
  \caption{Increase in the heating-related coefficient and the constant in the
    power consumption model for the different scenarios.}
  \label{tab:electrificationscenarios}
  \begin{tabular}{|l|r|r|r|r|}
    \hline
    \textbf{Country} & \multicolumn{2}{|c|}{\textbf{Half Electrification}} & \multicolumn{2}{|c|}{\textbf{Full Electrification}} \\
    \hline
    & Temp. Sens. Increase & Constant & Temp. Sens. Increase & Constant \\
    \hline
    Norway  &   8.7\% &  8.8 TWh &  17.4\% & 17.6 TWh \\
    Sweden  &  12.2\% & 12.7 TWh &  24.3\% & 25.4 TWh \\
    Denmark & 307.9\% &  6.0 TWh & 615.8\% & 11.9 TWh \\
    Finland &  64.5\% & 12.5 TWh & 129.0\% & 24.9 TWh \\
    \hline
  \end{tabular}
\end{table}

In order to explore the interaction between consumption and variable renewable
power generation, we also require assumptions regarding wind and solar power
generation capacities in 2040. When combined with weather data, these
assumptions will allow us to calculate residual demand, which is the power
consumption minus wind and solar power generation. We rely on the Nordic Clean
Energy Scenarios \citep{wrake_nordic_2021} for projections of the wind and solar
power generation capacities in 2040, which are shown in Table
\ref{tab:renewablescenarios}. When consumption and the generation in different
regions are aggregated, we assume for simplicity that there are no transmission
restrictions or losses (``copperplate transmission'').

\begin{table}
  \centering
  \caption{Assumed wind and PV capacities for 2040, from the Carbon Neutral
    Nordic (CNN) scenario of the Nordic Clean Energy Scenarios
    \citep{wrake_nordic_2021}.}
  \label{tab:renewablescenarios}
  \begin{tabular}{|l|r|r|}
    \hline
    \textbf{Country} & \textbf{Wind Generation Capacity 2040} & \textbf{PV
      Capacity 2040} \\
    \hline
    Norway  &  7.2 GW &  0.03 GW \\
    Sweden  & 21.8 GW &   7.1 GW \\
    Denmark & 20.0 GW &   9.1 GW \\
    Finland &  7.4 GW  &  7.5 GW \\
    \hline
  \end{tabular}
\end{table}

Comparing the projected probability distributions from the simulations of
partial or full heating electrification (HALF or FULL) to the simulations of no
further heating electrification (BAU) should illustrate clearly what effects
heating electrification will have on power consumption, and the residual demand
will show how variable renewable power generation will interact with the power
consumption.

\subsection{Weather Scenarios}
\label{sec:weatherscen}
For our strategy for investigating weather risk, we generate 300 simulated
weather paths using the shifted date method, which has been shown to produce
accurate probabilistic load forecasts \citep{xie_temperature_2018}. In the
shifted date method, we use entire historical weather years that are shifted
backwards or forwards by a certain number of days, which ensures that each
weather scenario is geographically and temporally consistent. Although this
method does not capture long-term changes that may be occurring in the climate,
\cite{van_der_wiel_influence_2019} have shown that changes due to climate change
are substantially smaller than interannual weather variability, which is the
main focus of our study. We then feed the weather paths into the consumption
models, together with population and GDP projections for 2040, in order to
create a large amount of consumption scenarios for 2040 at hourly resolution
under differing weather conditions.

Scenarios for variable renewable power generation -- wind and solar power -- are
coupled with the consumption scenarios, that is, based on the same weather
paths. The scenarios for wind and solar are based on the capacity factors used
by \cite{grams_balancing_2017}. Using consumption paths coupled with wind and
solar power generation based on the same weather conditions will show how power
consumption and VRE generation interact, and provide a deeper understanding of
how heating electrification will affect the weather risk of the Nordic power
system within the context of the large expected expansion in VRE generation
capacity.

\subsection{Data Sources and Preprocessing}
Hourly power consumption data have been provided by each countries' power grid
operator. Historical and projected GDP data have been retrieved from the
long-term real GDP forecast published by \cite{oecd_real_2018}, whereas
historical and forecasted population was retrieved from the OECD.Stat
database\footnote{Historical population available at
  \url{https://stats.oecd.org/Index.aspx?DataSetCode=HISTPOP}, and population
  projections at \url{https://stats.oecd.org/Index.aspx?DataSetCode=POPPROJ},
  accessed Oct. 20, 2021.}. The GDP and population data were transformed to
hourly data by linear interpolation.

Hourly weather data at five locations for each of the countries from 1985 to
2016 was retrieved from the ERA5 reanalysis data on single levels, provided by
\cite{hersbach_era5_2018} through the Copernicus Climate Change Service (C3S)
Climate Data Store (CDS). Reanalysis data are convenient for this study, as this
ensures that the data is complete and consistent.

Hourly capacity factors for wind and solar power generation were provided by
\cite{grams_balancing_2017}. To calculate the power generation projections, the
capacity factors were multiplied by their respective projections for installed
capacity.

\section{Results and Discussion}
\subsection{Consumption Model Estimation}
We calibrate the hourly power consumption models for each of the countries and
calculate accuracy metrics as discussed in Section \ref{sec:consmodels}. The
regression results are shown in Appendix \ref{appendix:a}, whereas the accuracy
metrics are shown in Table \ref{tab:errors}. Generally, the mean absolute
percentage error (MAPE) appears to be around 5\% for these countries, with only
small differences between the training sample, the validation sample and the
full sample. The models do not appear to suffer from overfitting, since the
performance on the training and validation samples are very close.

Overall, the accuracy of the models is reasonable for such simple models,
although a little lower than efforts with a greater focus on model accuracy
(such as \cite{mcsharry_probabilistic_2005, hor_analyzing_2005,
  hyndman_density_2010, trotter_climate_2016, rodriguez_climate_2019}), which
often achieve a MAPE of around 2\%-3\%. We believe it would be possible to
achieve a similar accuracy in our case, but this would introduce additional
model complexity which may not necessarily be appropriate for our purposes. For
the purposes of this study, we are satisfied with the accuracy offered by these
relatively simple models.

\begin{table}
  \centering
  \caption{Consumption model accuracy metrics. Ordinary least squares regression.}
  \label{tab:errors}
  \begin{tabular}{|l|r|r|r|r|}
    \hline
    \multicolumn{5}{|c|}{\textbf{Training Sample}} \\
    \hline
    \textbf{Country} & \textbf{RMSE} & \textbf{MAE} & \textbf{MAPE} &
    \textbf{sMAPE} \\
    \hline
    Norway &  760.3 MWh & 584.0 MWh & 4.2\% & 4.2\% \\
    Sweden &  862.8 MWh & 687.0 MWh & 4.4\% & 4.4\% \\
    Denmark & 276.6 MWh & 217.8 MWh & 5.6\% & 5.6\% \\
    Finland & 593.5 MWh & 429.5 MWh & 4.9\% & 4.8\% \\

    \hline
    \multicolumn{5}{|c|}{\textbf{Validation Sample}} \\
    \hline
    \textbf{Country} & \textbf{RMSE} & \textbf{MAE} & \textbf{MAPE} &
    \textbf{sMAPE} \\
    \hline
    Norway &  759.1 MWh & 582.7 MWh & 4.2\% & 4.2\% \\
    Sweden &  869.4 MWh & 691.8 MWh & 4.4\% & 4.4\% \\
    Denmark & 277.0 MWh & 218.7 MWh & 5.6\% & 5.6\% \\
    Finland & 599.9 MWh & 432.2 MWh & 5.0\% & 4.9\% \\

    \hline
    \multicolumn{5}{|c|}{\textbf{Full Sample}} \\
    \hline
    \textbf{Country} & \textbf{RMSE} & \textbf{MAE} & \textbf{MAPE} &
    \textbf{sMAPE} \\
    \hline
    Norway &  760.1 MWh & 583.8 MWh & 4.2\% & 4.2\% \\
    Sweden &  864.4 MWh & 688.2 MWh & 4.4\% & 4.4\% \\
    Denmark & 276.6 MWh & 218.0 MWh & 5.6\% & 5.6\% \\
    Finland & 595.2 MWh & 430.4 MWh & 4.9\% & 4.8\% \\
    \hline
  \end{tabular}
\end{table}

\begin{figure}
  \centering
  \begin{subfigure}[b]{0.48\textwidth}
    \centering
    \includegraphics[width=\textwidth]{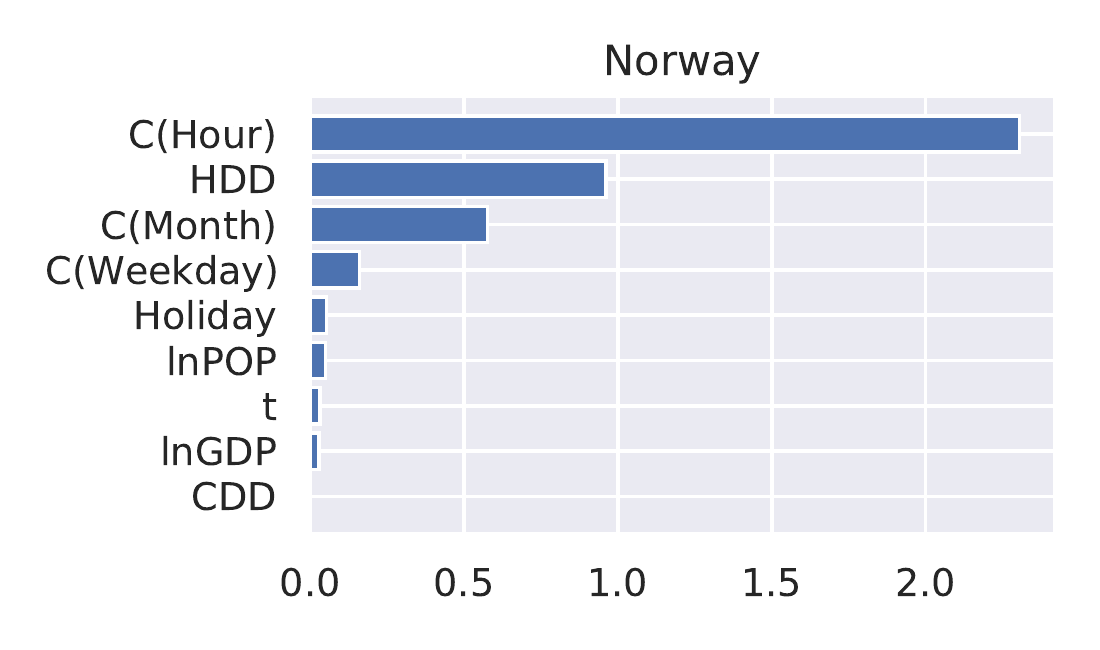}
    \label{fig:importanceNO}
  \end{subfigure}
  \hfill
  \begin{subfigure}[b]{0.48\textwidth}
    \centering
    \includegraphics[width=\textwidth]{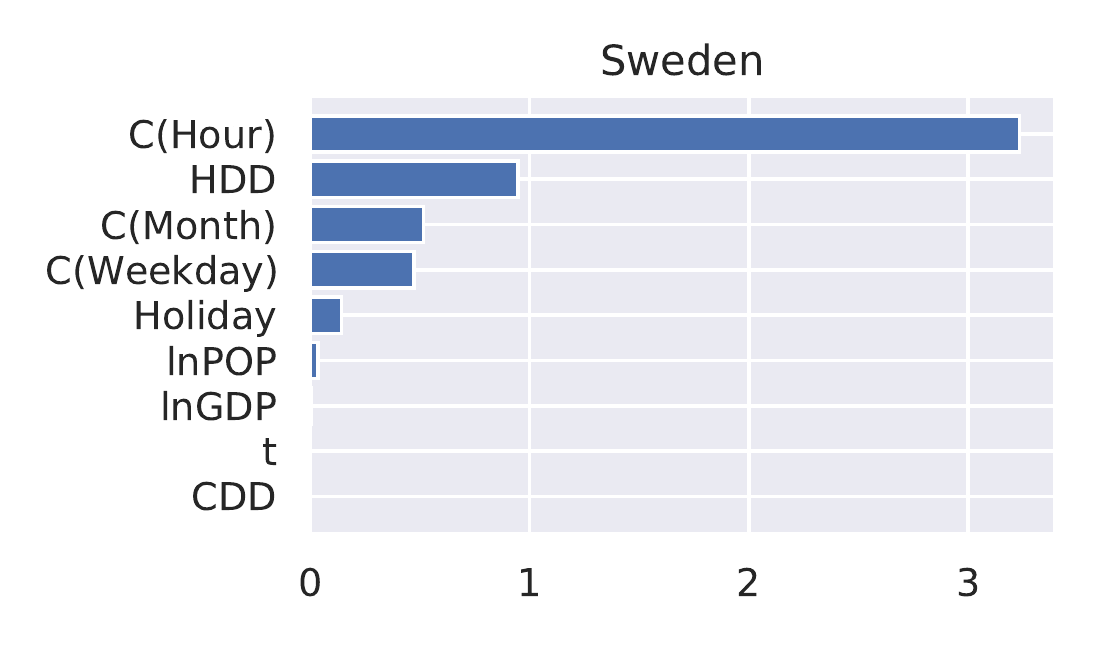}
    \label{fig:importanceSE}
  \end{subfigure}
  
  \begin{subfigure}[b]{0.48\textwidth}
    \centering
    \includegraphics[width=\textwidth]{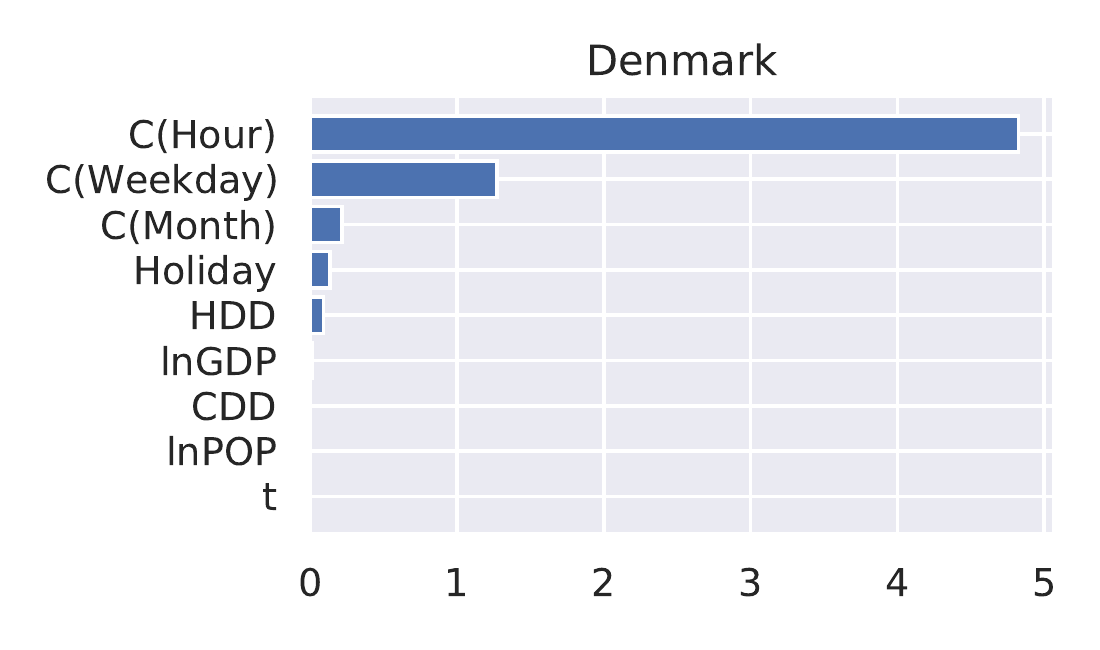}
    \label{fig:importanceDE}
  \end{subfigure}
  \hfill
  \begin{subfigure}[b]{0.48\textwidth}
    \centering
    \includegraphics[width=\textwidth]{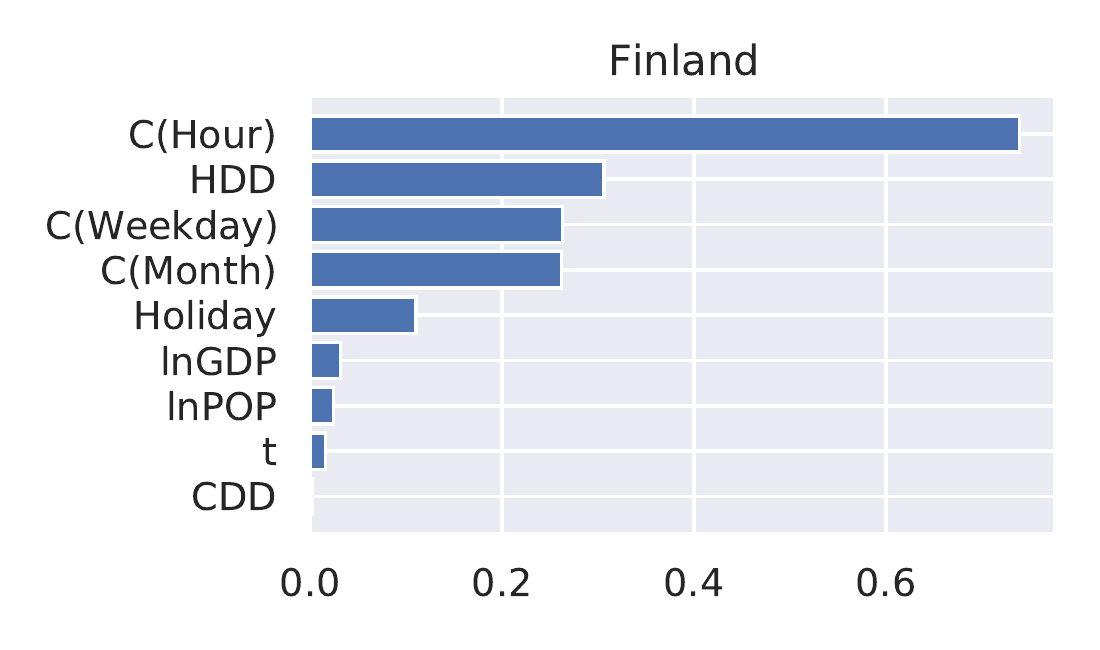}
    \label{fig:importanceFI}
  \end{subfigure}
  \caption{Effect size of the different model elements for the different
    countries, calculated by Cohen's $f^2$, \cite{selya_practical_2012}.}
  \label{fig:importance}
\end{figure}

In order to verify the importance of weather in the power consumption models, we
calculate the effect size of the different elements in the model, using Cohen's
$f^2$ \citep{selya_practical_2012}. The results are illustrated in Figure
\ref{fig:importance}, which shows that the importance of the HDH variables are
second only to the hourly profile for Norway, Sweden, and Finland. For Denmark,
in which much less of the heating is electric, the HDH variables are of lower
importance, as the hourly, weekday, monthly and holiday profiles have a higher
importance. However, since these seasonalities are entirely deterministic, this
nonetheless confirms that weather is the most important non-deterministic driver
of power consumption in the Nordic countries.

Therefore, we are satisfied that the consumption models that we have calibrated
are both capable and appropriate for the purposes of simulating increased
heating electrification in the Nordic region, and proceed to running the
simulations for year 2040, described in Sections \ref{sec:elecscen} and
\ref{sec:weatherscen}.

\subsection{Total Annual Consumption}

We now turn to the results from simulating the heating electrification
scenarios. For each of the three heating electrification scenarios, 300
simulations are run with different weather conditions, generated using the
shifted date method. From these simulations, we calculate key figures -- such as
total and peak consumption -- then examine and compare their distributions
between each of the heating electrification scenarios.

\begin{figure}
  \centering
  \includegraphics[width=\textwidth]{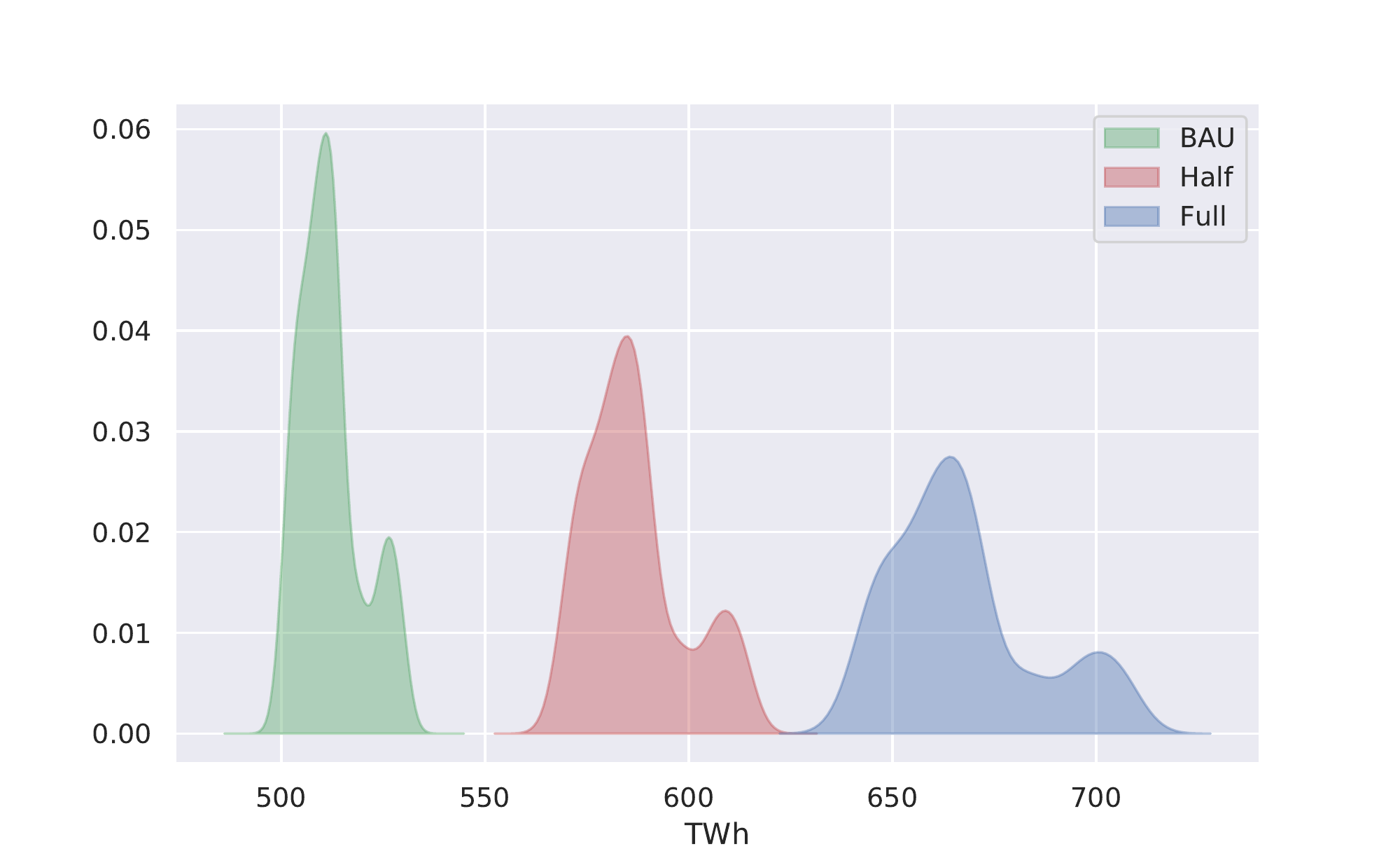}
  \caption{Projected densities of the projected total annual electricity
    consumption in the Nordic countries in 2040 for the three heating
    electrification scenarios (business-as-usual, half and full replacement of
    fossil fuels by electricity), under varying weather conditions.}
  \label{fig:totalcons}  
\end{figure}

Total consumption measures how much energy is consumed in the course of a year,
calculated by summing the power consumption over all hours of the year. The
distributions of total consumption for each of the three heating electrification
scenarios are illustrated in Figure \ref{fig:totalcons}, where a Gaussian kernel
has been used to create smooth density esimates. In the figure, heating
electrification can be seen to have two distinct effects on the total
consumption. Firstly, the electrification of heating increases the level of
total consumption, since the entire distributions for the half and full
electrification scenarios shift to the right. Secondly, the electrification of
heating also widens the distributions for total consumption considerably. This
shows that heating electrification not only causes increased power consumption,
but also substantially higher weather-based variation in annual power
consumption\footnote{The impacts on individual countries in the region are
  similar, as shown in Appendix \ref{appendix:b}, all of which show both a right
  shift and a widening of the probability distributions with increased heating
  electrification.}.

In order to explicitly quantify the impacts of heating electrification, we
calculate the mean, standard deviation, and conditional value-at-risk at the 5\%
level (CVaR$_{5\%}$) of the distribution of total annual consumption over all
the weather scenarios for each of the heating electrification scenarios. In this
context, the CVaR$_{5\%}$ can be interpreted as a typical one-in-twenty
occurrence. Table \ref{tab:resultssummary} shows these key indicators.

The mean of the total consumption in the full electrification scenario is around
155 TWh (30\%) higher than the business-as-usual (BAU) scenario. The
CVaR$_{5\%}$ of total consumption, however, is almost 178 TWh (34\%) higher in
the full electrification scenario compared to the BAU scenario. This shows that
full heating electrification can cause a large increase in total power
consumption in a normal year, but an even greater increase during an unusually
cold year.
\begin{table}
  \centering
  \caption{Summary of the simulation results.}
  \label{tab:resultssummary}
  \begin{tabular}{|l|r|r|r|}
    \hline
    \multicolumn{4}{|c|}{\textbf{Total Consumption (TWh)}} \\
    \hline
    \textbf{Scenario} & \textbf{Mean} & \textbf{Std. dev.} &
    \textbf{CVaR}$_{5\%}$ \\
    \hline
    Business As Usual     & 512.1 &  7.7 & 528.9  \\
    Half Electrification  & 586.7 & 11.9 & 612.9  \\
    Full Electrification  & 667.2 & 17.7 & 706.6  \\
    \hline
    \multicolumn{4}{|c|}{\textbf{Peak Hour Consumption (GWh)}} \\
    \hline
    \textbf{Scenario} & \textbf{Mean} & \textbf{Std. dev.} &
    \textbf{CVaR}$_{5\%}$ \\
    \hline
    Business As Usual     &  88.6 &  5.0 & 102.8  \\
    Half Electrification  & 108.8 &  8.4 & 133.0  \\
    Full Electrification  & 134.8 & 14.3 & 176.6  \\
    \hline
    \multicolumn{4}{|c|}{\textbf{Peak Hour Residual Demand (GWh)}} \\
    \hline
    \textbf{Scenario} & \textbf{Mean} & \textbf{Std. dev.} &
    \textbf{CVaR}$_{5\%}$ \\
    \hline
    Business As Usual     &  79.7 &  5.0 &  92.5  \\
    Half Electrification  &  98.9 &  8.1 & 118.4  \\
    Full Electrification  & 123.8 & 13.4 & 154.8  \\
    \hline
  \end{tabular}
\end{table}

In comparison, the Nordic Clean Energy Scenarios (NCES) developed by
\cite{wrake_nordic_2021} project a total consumption between 378 and 423 TWh in
year 2040, as the Nordic region transitions to carbon neutrality in year
2050. These projections are even below our business-as-usual scenario, even
though they incorporate consumption increases from additional sectors, such as
transport and datacenters. The main reason that the NCES project lower total
consumption, appears to be that they explicitly assume a greater increase in
efficiency, and a transition to non-electric district heating (for instance
waste). The carbon neutral scenario of \cite{iea_nordic_2016} expects a total
consumption of around 375 TWh in year 2040, which is also below our projections
as they assume a large increase in the efficiency of electric heating. Our
scenarios, however, do not make explicit assumptions regarding increased
efficiency of the existing electric heating, only for the heat pumps that
replace fossil-based heating. However, a study by \cite{halvorsen_hvem_2013}
claims that Norwegian consumers entirely offset the saved energy by increased
consumption when replacing resistive heating with heat pumps, such that the
increased efficiency does not lead to energy savings. As such, it is not certain
that increased efficiency will lead to large drops in total consumption. If this
turns out to be the case, the consumption projections by both
\cite{iea_nordic_2016} and \cite{wrake_nordic_2021} will turn out to be too low,
and, furthermore, our electrification scenarios might also underestimate total
consumption.

The long-term outlook report by the Norwegian power grid operator Statnett
\citep{statnett_langsiktig_2020}, however, projects a Nordic consumption of 579
TWh in year 2040, which also includes contributions from the transport sector
and datacenters, as well as increased electrification in other
sectors. Discounting the transport and datacenter sectors, consumption is
projected at 479 TWh. This is relatively close to the mean of our
business-as-usual scenario at 512.1 TWh, but significantly below our scenarios
with heating electrification -- especially for a one-in-twenty year.

Counting both electrification in existing sectors and new power consuming
sectors such as transportation, datacenters, and hydrogen, The Norwegian Water
Resources and Energy Directorate (NVE) projected a total Nordic consumption of
about 480 TWh in year 2040 in their 2020 report
\citep{noregs_vassdrags-_og_energidirektorat_nve_langsiktig_2020}, but revised
their projections to 526 TWh in their 2021 report
\citep{noregs_vassdrags-_og_energidirektorat_nve_langsiktig_2021}. These figures
are relatively close to our business-as-usual scenario. However, they include
contributions from new sectors that have not been explicitly accounted for in
our projections, which simply assume a continuation general trend that has been
observed in the last few decades. Even though the upcoming changes projected by
NVE appear to imply greater consumption increases in the coming decades than has
been experienced in the last few decades, our projections still appear to be
higher than the projections by NVE.

Therefore, our results imply a total consumption that is considerably higher
than previous estimates, especially in the scenarios with heating
electrification -- and even more so in a cold year: for a one-in-twenty year in
the full electrification scenario, total consumption is almost 90\% above the
lowest scenarios by \cite{iea_nordic_2016}, and still more than 45\% higher than
the comparable projection by \cite{statnett_langsiktig_2020}. In that sense, it
is possible that heating electrification may present significantly larger
challenges for the Nordic power system than previously believed.

\subsection{Peak Consumption and Peak Residual Demand}
\begin{figure}
  \centering
  \includegraphics[width=\textwidth]{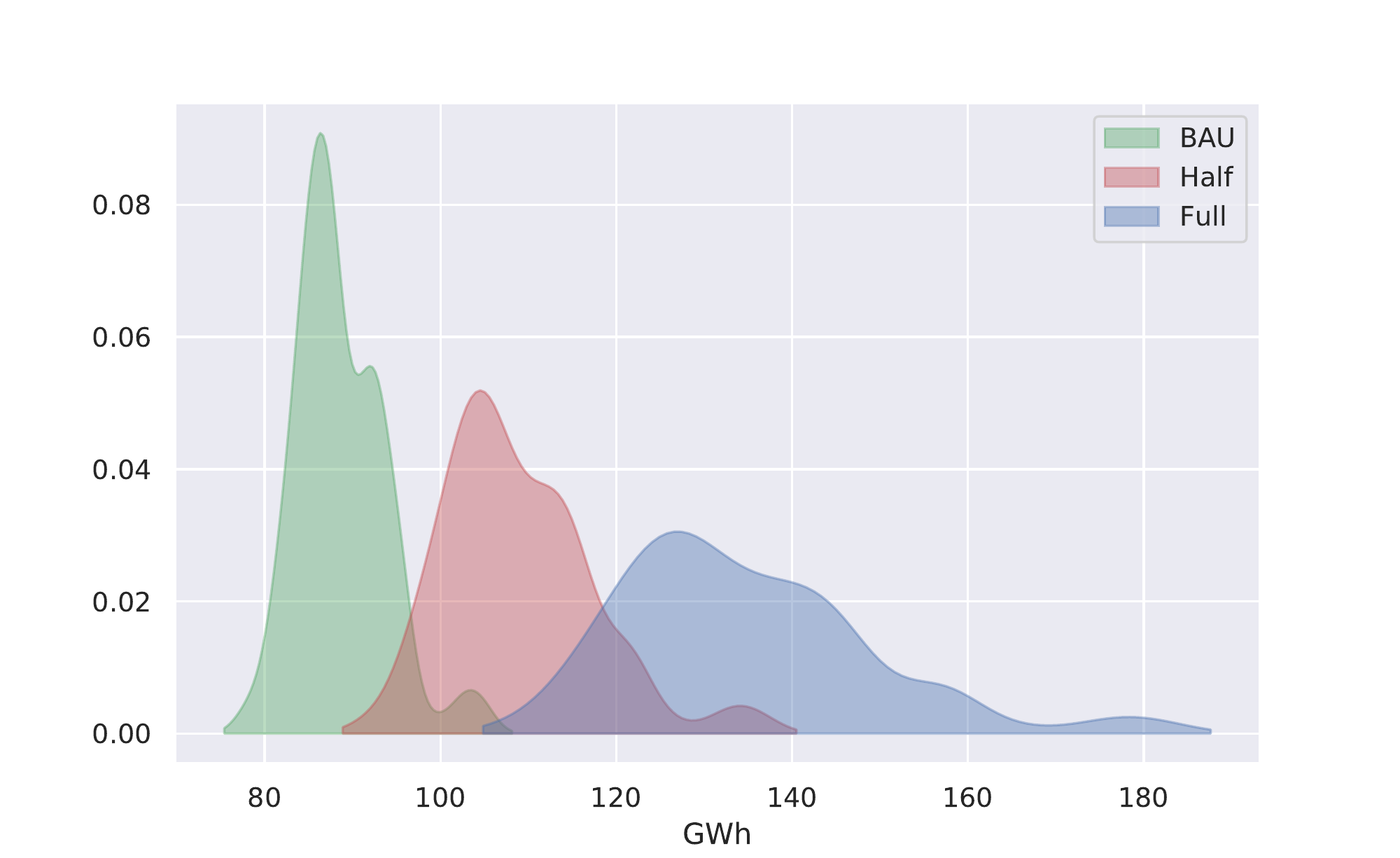}
  \caption{Projected densities of the Nordic electricity consumption in the peak
    hour in 2040 for the three heating electrification scenarios
    (business-as-usual, half and full replacement of fossil fuels by
    electricity), under varying weather conditions.}
  \label{fig:peakcons}  
\end{figure}

Peak consumption is a key figure for planning purposes, since the power system
must be designed to withstand peak load. The two central features we observed
with total consumption are also present in peak consumption, illustrated in
Figure \ref{fig:peakcons}: heating electrification shifts the distribution to
the right, and at the same time widens the distribution. Full electrification
increases the consumption in the peak hour in the Nordic countries from 88.6 GWh
in the BAU scenario to 134.8 GWh in the full electrification scenario, an
increase of about 52\%, for a normal weather year. For a one-in-twenty year,
however, the full electrification scenario is almost 72\% above the BAU
scenario. This shows that the effects of electrification are even more serious
for peak consumption than for total consumption.

In comparison, \cite{statnett_langsiktig_2020} project a peak load of 95 GW in
year 2040, which is close to the mean of our business-as-usual scenario, but
significantly lower than the electrification scenarios and our estimates for a
typical one-in-twenty weather year. This also implies that the challenges posed
by heating electrification may be greater than previously believed.

Comparing the peak consumption to installed power generation capacity is a
simple and intuitive way to check if it will be possible for the power system to
supply the peak load. \cite{wrake_nordic_2021} project an installed power
generation capacity between 142 and 161 GW in year 2040, which at first sight
appears to be adequate for all but a one-in-twenty weather year with full
heating electrification. Generation capacity projections by
\cite{statnett_langsiktig_2020} are slightly higher, at 175 GW in year 2040.

However, some of the power generation also depends directly on weather
conditions and is not dispatchable -- it is not certain that this capacity can
be relied on to produce sufficient power to supply the peak. In the projections
by \cite{wrake_nordic_2021}, installed capacity drops to about 70 GW if we
exclude solar and wind power. Excluding non-dispatchable capacity from the
projections of \cite{statnett_langsiktig_2020} gives a generation capacity of
around 68 GW.

\begin{figure}
  \centering
  \includegraphics[width=\textwidth]{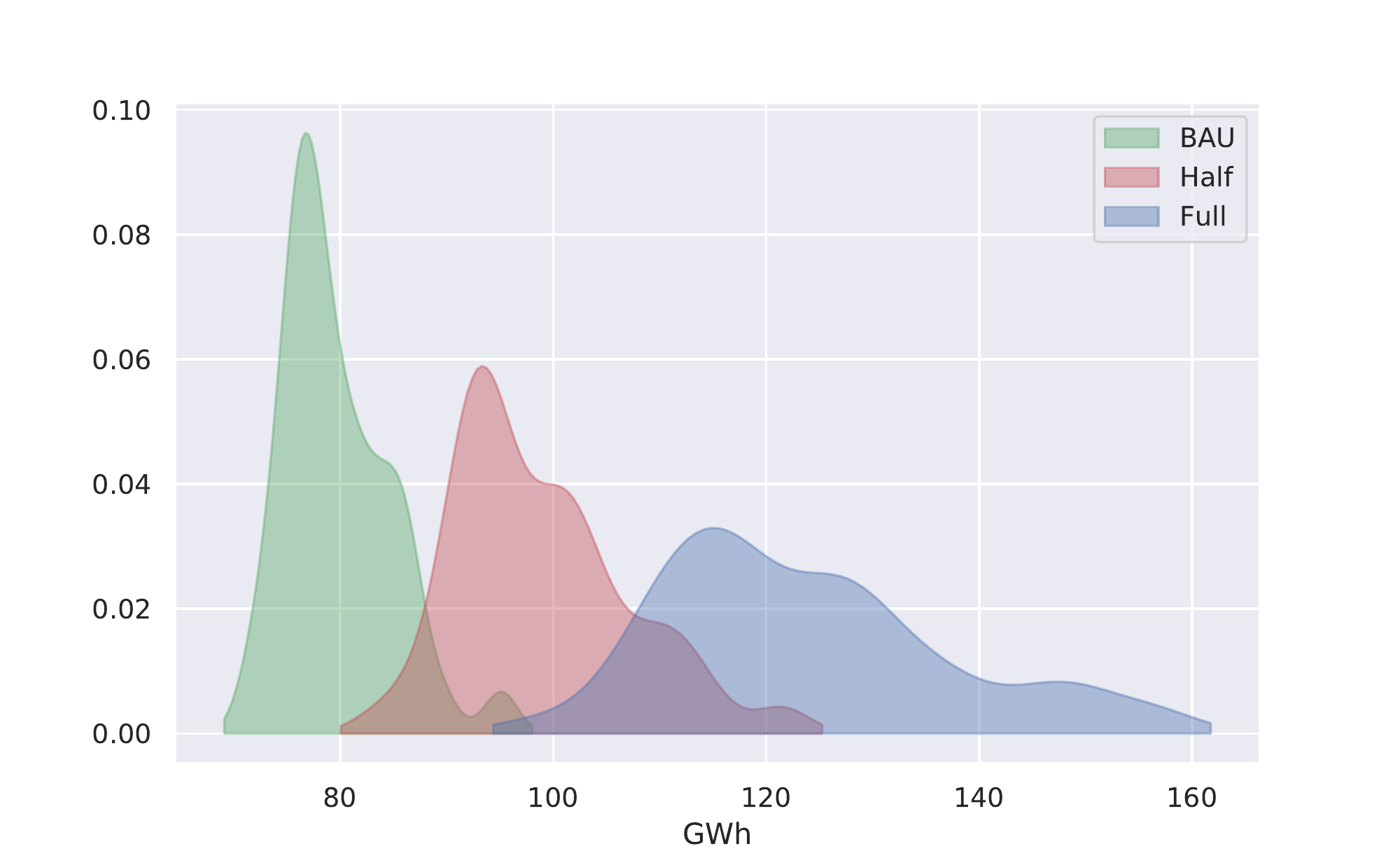}
  \caption{Projected densities of the Nordic residual electricity demand in the
    peak hour in 2040 for the three electrification scenarios
    (business-as-usual, half and full replacement of fossil fuels by
    electricity), under varying weather conditions.}
  \label{fig:peakresdem}
\end{figure}

With this in mind, we calculate the residual demand, which is the power
consumption minus solar and wind power generation. The residual demand therefore
gives an idea of how much energy must be produced by generation sources apart
from the variable renewable sources. The distribution of the residual demand in
the peak hour is shown in Figure \ref{fig:peakresdem}, and some summary
statistics are shown in Table \ref{tab:resultssummary}. The figure shows that
the peak residual demand shares the two main features of total and peak
consumption: heating electrification causes the distribution to shift to the
right, and the distribution becomes significantly wider -- higher levels and
increased risk. For all three electrification scenarios, the mean of the peak
hour residual demand is about 10 GWh lower than the corresponding peak hour
consumption. That implies that when they are needed the most -- in the peak
consumption hour -- variable renewable sources are only providing about
12\%-13\% of their nameplate capacity. The residual demand must then be supplied
by power generation from other sources. It therefore appears that the projected
non-VRE generation capacity of around 70 GW would be unable to supply the
residual demand during the peak hour of a typical weather year for none of the
three electrification scenarios -- business-as-usual (79.7 GWh), half
electrification (98.9 GWh), or full electrification (123.8 GWh).

In addition, when calculating the aggregate peak residual demand in the entire
Nordic region, we have assumed that there are no limitations or losses related
to transmission within the region (``copperplate transmission''). There could be
greater variation at the local level, such that the situations at the local
level would likely be more serious if one were to take transmission constraints
and losses into account.

Since projected non-VRE generation capacity would be insufficient to supply the
residual demand during the peak hour, the Nordic countries could perhaps import
power from other countries during these critical
periods. \cite{wrake_nordic_2021} project an import capacity around 20 GW,
bringing the non-VRE supply capacity to around 90 GW, implying that the system
could in principle rely on imports from other countries -- but only during a
fairly typical weather year with lower levels of heating electrification. With
full heating electrification, a one-in-twenty weather year would require 154.8
GWh during the peak hour, which exceeds the Nordic non-VRE supply capacity
projected by \cite{wrake_nordic_2021}. This means that the Nordic power system
might struggle to ensure a stable and secure power supply during peak hours
during a normal year with no further heating electrification, and could face
serious shortages at higher levels of heating electrification and during colder
weather years.

Therefore, our results suggest that increased electrification of heating in the
Nordic countries would require a substantial expansion in the power system
compared to what is expected in earlier studies. In particular, the increase in
weather risk caused by heating electrification implies that special attention
should be paid to ensuring sufficient flexibility to meet peak power demand in
periods with little wind or solar power generation.

\subsection{Load Duration}
\begin{figure}
  \centering
  \includegraphics[width=\textwidth]{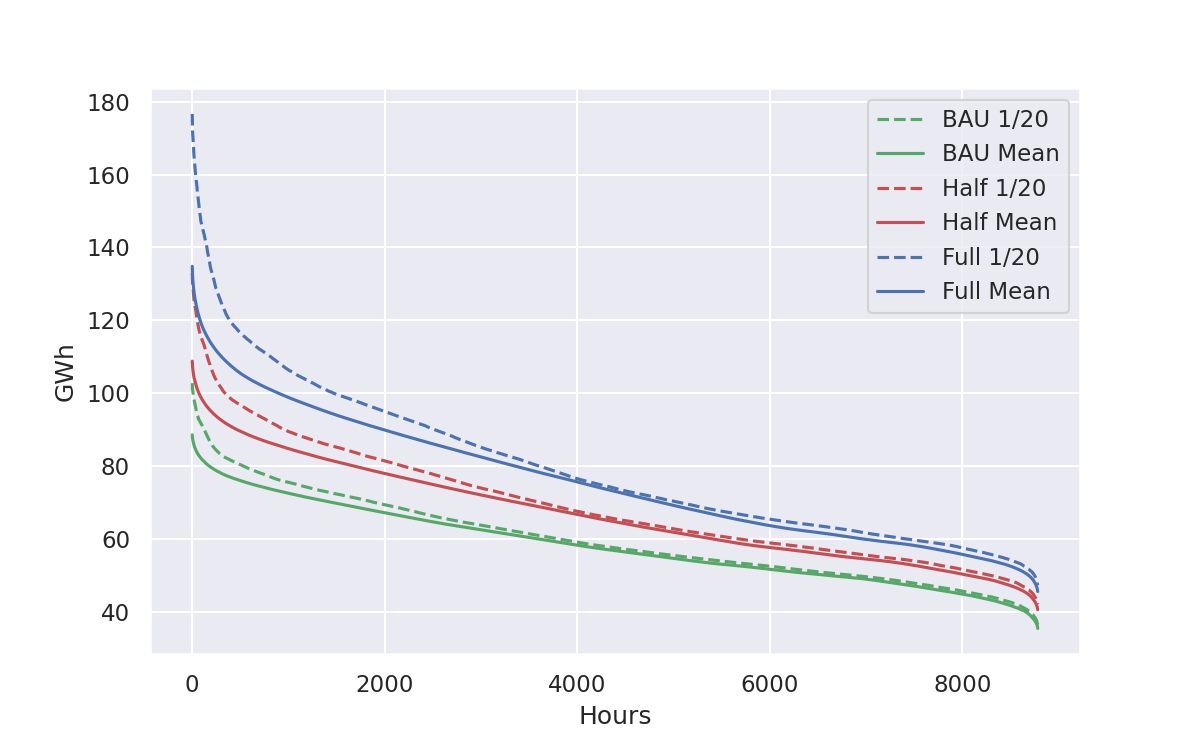}
  \caption{Projected load duration curves for 2040, for a typical year (Mean)
    and a one-in-twenty year (1/20) for the three heating electrification
    scenarios (business-as-usual, half and full replacement of fossil fuels by
    electricity).}
  \label{fig:load_duration}
\end{figure}

Load duration curves are an essential power system planning instrument, which
illustrates the duration at which the power system load is at or above a certain
level. Figure \ref{fig:load_duration} shows the load duraction curves for a
typical year and a typical one-in-twenty year, for each of the three heating
electrification scenarios.

We can see from the figure that heating electrification causes the load duration
curves to become much steeper, which implies a far greater variability in the
consumption. The increase in steepness is much more pronounced for the typical
one-in-twenty year.

Furthermore, the figure illustrates another serious aspect of the impacts of
heating electrification: the large increase in consumption that we observed in
the total annual consumption is in fact very disproportionally allocated
throughout the year: the periods with already high consumption also receive the
largest increase in consumption. As such, the periods that already experience
the highest consumption will experience even higher consumption, whereas
low-consumption periods will experience little change.

This disproportional impact means that the power system will not only need to be
designed to serve a higher peak consumption, but that the power system must be
designed to withstand larger seasonal differences, longer continous periods of
high consumption, and larger consumption variations at every timescale. As such,
the insights from this study may have large consequences for the design of the
future Nordic power system.

\subsection{Discussion}
In our results, we have highlighted the increase in the level, variability, and
uncertainty in Nordic power consumption caused by heating electrification, and
we have compared our results to existing projections for the Nordic power
system. This has revealed that existing projections might have underestimated
the impact that heating electrification would have on the Nordic power system,
and suggests that heating electrification must be accompanied by large increases
in power system flexibility at every timescale. Qualitatively, these findings
mirror many of the conclusions in the existing scientific literature.

By and large, our results are consistent with the findings of
\cite{wilson_historical_2013}, who analysed two years of historical gas and
power consumption data to show that migrating even a small proportion of natural
gas heating to power in Britain would cause a large increase in the level,
variability, and uncertainty of the power
consumption. \cite{staffell_increasing_2018} also support this conclusion in
their projections of UK power consumption in year 2030, showing increased
year-to-year, seasonal and intra-seasonal variability due to heating
electrification. Our results show that electrifying heating would have similar
impacts in the Nordic countries. \cite{quiggin_implications_2016} further imply
that several official projections for the UK do not adequately account for the
effects of heating electrification, which may cause balancing issues for the UK
power system in year 2050. Although \cite{watson_decarbonising_2019} believe
that \cite{quiggin_implications_2016} overestimate the peaks, they still agree
on the main aspects of how heating electrification will change power consumption
in the UK. Similar to \cite{quiggin_implications_2016}, our results suggest that
the serious impacts of heating electrification on power consumption may be
underestimated in the projections for the Nordic power system in year 2040.

In a study of the power system in Ireland in 2030 with increased electrical
heating and VRE share, \cite{heinen_electrification_2017} find that weather
would cause considerable variation in system costs and that cold and windless
weather define the most critical periods, which is consistent with our
results. The authors further suggest that pre-heating and the thermal inertia of
buildings could lower the power consumption peaks and, consequently, system
costs. We agree with this point, although our results show that heating
electrification not only increases the peaks, but also increases power
consumption for longer durations, requiring flexibility sources beyond the short
term alleviation provided by thermal inertia.

Both \cite{kirkerud_power--heat_2017} and \cite{ruhnau_heating_2020} find that
increased use of electricity for heating raises the wind market value,
considering district heating in Northern Europe in the former study and heat
pumps in the UK in the latter. However, the results in both studies imply that
the wind-driven electric heating is to a large degree mediated by storage, and
not driven directly by wind power generation. In addition, when
\cite{chen_impact_2021} found that electric heat pumps combined with wind power
expansion would be a promising decarbonisation strategy for Northern Europe,
they also pointed out that such a solution would require large increases in
power system flexibility. In that light, it is also not surprising that
\cite{haghi_iterative_2020} found that heating electrification combined with an
expansion of natural gas power generation would be a cost-effective solution for
reducing emissions in the UK, since natural gas generation is normally
relatively flexible compared to wind power. \cite{thomasen_decarbonisation_2021}
argue that some EU27+UK countries -- mainly with warm climates -- may already be
well prepared for full electrification of heating, but warn that many others --
often characterised by colder climates -- currently have insufficient firm
generation capacity for large-scale heating electrification. On the whole, the
results from these studies are consistent with the conclusions of our study: we
have shown that the wind power output in the Nordic countries is relatively low
during the most critical periods, which means that substantial increases in
power system flexibility would be required.

Therefore, our qualitative findings are well established within the context of
the existing scientific literature on heating electrification -- even though we
have arrived at them by different means and for a region that has not previously
been explicitly considered. However, comparing our results with existing
projections for the Nordic power system shows that the effects we have
identified do not appear to be appropriately incorporated, and therefore may not
haven been sufficiently well established until now. Furthermore, beyond the
qualitative conclusions we have reached, the magnitudes of the effects that we
have presented are also interesting and relevant to planners and policymakers,
and provide a deeper understanding of how heating electrification and VRE
expansion will impact weather risk in the Nordic power system.

\subsection{Study Caveats}
Given our stylised approach and the fact that we consider a year that is nearly
two decades in the future, it is important that the results of this study are
interpreted correctly, and therefore we discuss some of the main caveats of the
study.

Firstly, the structure of power consumption may change in the future compared to
the calibration period of the consumption models.
The impacts of the various model elements are not fixed in reality, but change
over time in response to the underlying technological and social conditions.
Policies affecting the calibration period, such as the Norwegian ban on mineral
oil for heating in 2020, might not impact year 2040 -- and policies that we have
not yet imagined may have a large impact in 2040.
Technological change may also cause large differences between the calibration
period and year 2040.
For instance, we have not explicitly considered how improvements in buildings'
thermal efficiency may affect our results.
Until 2040, older buildings with low thermal efficiency will be replaced with
new buildings, we might see increased adoption of heat pumps, and further
improvements in building techniques -- all of which will lower power consumption
for heating, and is not explicitly considered in the historical/statistical
power consumption models.
However, our power consumption models do contain a trend variable, so this is
only relevant if these changes occur at a different rate than in the calibration
period.
New power consuming sectors such as transportation, datacenters, hydrogen
production, and so forth, are also not explicitly represented in our
approach.
On the one hand, these sectors may come to represent a large proportion of power
consumption.
On the other hand, there is little historical data available to directly make
quantitative estimations about the impacts of these developments.
As such, we have refrained from making explicit assumptions about such
structural developments in order to focus exclusively on heating
electrification.

We may also see greater adoption of load-shifting and demand response
technologies in the future, which may reduce consumption in the peak
hours. Although this could contribute greatly to balancing the system in
critical periods, such technologies are not necessarily capable of contributing
during sustained periods of high consumption and low VRE production. Such
periods would remain challenging, and further research is needed on how to
address these challenges.

Secondly, we suspect that the resampling method we used for generating the
weather scenarios may not be very well-suited for analysing extreme events,
since few or no extreme events of a particular type may have occurred in the
historical dataset. Although \cite{xie_temperature_2018} have established that
the shifted date method is a good choice for load forecasting purposes, we
suspect that simulated weather conditions from numerical weather prediction
models -- such as those used by \cite{van_der_wiel_meteorological_2019} -- might
actually provide a more comprehensive foundation for risk assessments, although
at the cost of greater complexity.

Thirdly, the possible impacts of climatic change are outside the scope of our
study, but if there are increases in temperature, then heating needs would
naturally decrease. \cite{van_der_wiel_meteorological_2019} have concluded,
however, that interannual weather variability far exceed the impacts of climate
change, and we therefore believe that this consideration would not substantially
alter our results.

Finally, generation cost or other system implications have also been outside the
scope of this study. This has allowed us to simplify the consumption models and
the renewable generation capacity projections greatly -- as if they were
exogenous to the power market. However, in reality both consumption and capacity
investments are endogenously determined in an interaction with a complex energy
market and an even more complex political and social environment. In particular,
persistent high or low prices will affect consumer behaviour, the development
and adoption of different technologies, or investments in different types of
generation capacity.

Although the simplifications we have made affect the results of our experiment,
they do not affect the main qualitative conclusions we have drawn: increased
heating electrification in the Nordic countries will lead to a relatively large
increase in total power consumption, and an even larger increase in weather risk
due to the increased weather sensitivity, whereas variable renewable sources
seem to contribute little to mitigating the risk. Nor do we believe that the
simplifications invalidate the additional implications we have inferred from our
experiment: if heating is to be electrified, the Nordic power system must not
only be designed to supply a greater total load, but also to endure much larger
inter-annual, seasonal and intra-seasonal variations.

\section{Conclusion}
We have shown that replacing heating based on fossil fuels with electric heating
in the Nordic countries will result in a considerably higher total power
consumption in 2040. More importantly, however, we have shown that it will lead
to greater inter-annual, seasonal and intra-seasonal variability in power
consumption, and the increased sensitivity to weather conditions resulting from
the electrification implies an increased exposure weather risk. Compared to
existing power system projections, our calculations suggest that both the
magnitude of the consumption increase caused by heating electrification and the
increase in weather risk may previously have been underestimated. As such, a
large expansion in power system flexibility will be necessary if heating is
electrified in the Nordic countries. Although variable renewable generation
capacity is projected to increase substantially in the future, these sources
appear to contribute little during the periods of the highest consumption, which
underlines the importance of ensuring sufficient flexibility in the power
system.

Given these considerations, we suggest that future research focuses on
determining how the Nordic power system can be developed to provide sufficient
flexbility to handle the changes we have identified. Such an analysis would
ideally consider multiple competing sources of flexibility, both on the demand
side and the supply side, and use techniques that could appropriately account
for variability, uncertainty, and contingency in a realistic manner.

This research is fundamental for designing a robust and resilient energy system
that accounts for the increased weather risk in power consumption and
production, while at the same time contributing to reduced greenhouse gas
emissions.



\bibliographystyle{elsarticle-harv}
\bibliography{Weather_Risk.bib}

\newpage

\appendix

\section{Appendix: Consumption Model Regression Results}
\label{appendix:a}
\begin{table}[!htbp] \centering
\begin{adjustbox}{totalheight=\textheight-3\baselineskip}  
\begin{tabular}{@{\extracolsep{5pt}}lcccc}
\\[-1.8ex]\hline
\hline \\[-1.8ex]
& \multicolumn{4}{c}{\textit{Dependent variable:} $\ln(\mathrm{Cons}_t)$} \
\cr \cline{2-5}
\\[-1.8ex] & Norway & Sweden & Denmark & Finland \\
\hline \\[-1.8ex]
 C(Hour)[T.10] & 0.166$^{***}$ & 0.247$^{***}$ & 0.393$^{***}$ & 0.135$^{***}$ \\
  & (0.001) & (0.001) & (0.001) & (0.001) \\
 C(Hour)[T.11] & 0.177$^{***}$ & 0.256$^{***}$ & 0.393$^{***}$ & 0.138$^{***}$ \\
  & (0.001) & (0.001) & (0.001) & (0.001) \\
 C(Hour)[T.12] & 0.179$^{***}$ & 0.255$^{***}$ & 0.375$^{***}$ & 0.134$^{***}$ \\
  & (0.001) & (0.001) & (0.001) & (0.001) \\
 C(Hour)[T.13] & 0.176$^{***}$ & 0.246$^{***}$ & 0.368$^{***}$ & 0.128$^{***}$ \\
  & (0.001) & (0.001) & (0.001) & (0.001) \\
 C(Hour)[T.14] & 0.173$^{***}$ & 0.239$^{***}$ & 0.351$^{***}$ & 0.129$^{***}$ \\
  & (0.001) & (0.001) & (0.001) & (0.001) \\
 C(Hour)[T.15] & 0.171$^{***}$ & 0.239$^{***}$ & 0.325$^{***}$ & 0.130$^{***}$ \\
  & (0.001) & (0.001) & (0.001) & (0.001) \\
 C(Hour)[T.16] & 0.168$^{***}$ & 0.235$^{***}$ & 0.337$^{***}$ & 0.136$^{***}$ \\
  & (0.001) & (0.001) & (0.001) & (0.001) \\
 C(Hour)[T.17] & 0.164$^{***}$ & 0.240$^{***}$ & 0.401$^{***}$ & 0.139$^{***}$ \\
  & (0.001) & (0.001) & (0.001) & (0.001) \\
 C(Hour)[T.18] & 0.159$^{***}$ & 0.231$^{***}$ & 0.393$^{***}$ & 0.145$^{***}$ \\
  & (0.001) & (0.001) & (0.001) & (0.001) \\
 C(Hour)[T.19] & 0.151$^{***}$ & 0.212$^{***}$ & 0.339$^{***}$ & 0.137$^{***}$ \\
  & (0.001) & (0.001) & (0.001) & (0.001) \\
 C(Hour)[T.1] & -0.032$^{***}$ & -0.025$^{***}$ & -0.050$^{***}$ & -0.030$^{***}$ \\
  & (0.001) & (0.001) & (0.001) & (0.001) \\
 C(Hour)[T.20] & 0.138$^{***}$ & 0.190$^{***}$ & 0.283$^{***}$ & 0.107$^{***}$ \\
  & (0.001) & (0.001) & (0.001) & (0.001) \\
 C(Hour)[T.21] & 0.120$^{***}$ & 0.160$^{***}$ & 0.237$^{***}$ & 0.113$^{***}$ \\
  & (0.001) & (0.001) & (0.001) & (0.001) \\
 C(Hour)[T.22] & 0.089$^{***}$ & 0.111$^{***}$ & 0.171$^{***}$ & 0.098$^{***}$ \\
  & (0.001) & (0.001) & (0.001) & (0.001) \\
 C(Hour)[T.23] & 0.045$^{***}$ & 0.052$^{***}$ & 0.081$^{***}$ & 0.047$^{***}$ \\
  & (0.001) & (0.001) & (0.001) & (0.001) \\
 C(Hour)[T.2] & -0.051$^{***}$ & -0.036$^{***}$ & -0.074$^{***}$ & -0.045$^{***}$ \\
  & (0.001) & (0.001) & (0.001) & (0.001) \\
 C(Hour)[T.3] & -0.059$^{***}$ & -0.041$^{***}$ & -0.079$^{***}$ & -0.051$^{***}$ \\
  & (0.001) & (0.001) & (0.001) & (0.001) \\
 C(Hour)[T.4] & -0.060$^{***}$ & -0.040$^{***}$ & -0.070$^{***}$ & -0.035$^{***}$ \\
  & (0.001) & (0.001) & (0.001) & (0.001) \\
 C(Hour)[T.5] & -0.044$^{***}$ & -0.009$^{***}$ & -0.020$^{***}$ & 0.031$^{***}$ \\
  & (0.001) & (0.001) & (0.001) & (0.001) \\
 C(Hour)[T.6] & 0.002$^{**}$ & 0.073$^{***}$ & 0.110$^{***}$ & 0.077$^{***}$ \\
  & (0.001) & (0.001) & (0.001) & (0.001) \\
 C(Hour)[T.7] & 0.075$^{***}$ & 0.162$^{***}$ & 0.260$^{***}$ & 0.106$^{***}$ \\
  & (0.001) & (0.001) & (0.001) & (0.001) \\
 C(Hour)[T.8] & 0.125$^{***}$ & 0.201$^{***}$ & 0.337$^{***}$ & 0.119$^{***}$ \\
  & (0.001) & (0.001) & (0.001) & (0.001) \\
 C(Hour)[T.9] & 0.149$^{***}$ & 0.223$^{***}$ & 0.369$^{***}$ & 0.126$^{***}$ \\
  & (0.001) & (0.001) & (0.001) & (0.001) \\
 C(Month)[T.10] & -0.105$^{***}$ & -0.084$^{***}$ & -0.072$^{***}$ & -0.076$^{***}$ \\
  & (0.001) & (0.001) & (0.001) & (0.001) \\
 C(Month)[T.11] & -0.035$^{***}$ & -0.037$^{***}$ & -0.026$^{***}$ & -0.035$^{***}$ \\
  & (0.001) & (0.001) & (0.001) & (0.001) \\
 C(Month)[T.12] & -0.013$^{***}$ & -0.016$^{***}$ & -0.023$^{***}$ & -0.044$^{***}$ \\
  & (0.001) & (0.001) & (0.001) & (0.001) \\
 C(Month)[T.2] & -0.006$^{***}$ & -0.006$^{***}$ & -0.012$^{***}$ & -0.010$^{***}$ \\
  & (0.001) & (0.001) & (0.001) & (0.001) \\
 C(Month)[T.3] & -0.036$^{***}$ & -0.046$^{***}$ & -0.052$^{***}$ & -0.047$^{***}$ \\
  & (0.001) & (0.001) & (0.001) & (0.001) \\
 C(Month)[T.4] & -0.113$^{***}$ & -0.108$^{***}$ & -0.107$^{***}$ & -0.098$^{***}$ \\
  & (0.001) & (0.001) & (0.001) & (0.001) \\
 C(Month)[T.5] & -0.178$^{***}$ & -0.155$^{***}$ & -0.130$^{***}$ & -0.144$^{***}$ \\
  & (0.001) & (0.001) & (0.001) & (0.001) \\
 C(Month)[T.6] & -0.211$^{***}$ & -0.171$^{***}$ & -0.122$^{***}$ & -0.218$^{***}$ \\
  & (0.001) & (0.001) & (0.001) & (0.001) \\
 C(Month)[T.7] & -0.258$^{***}$ & -0.254$^{***}$ & -0.184$^{***}$ & -0.170$^{***}$ \\
  & (0.001) & (0.001) & (0.002) & (0.001) \\
 C(Month)[T.8] & -0.213$^{***}$ & -0.171$^{***}$ & -0.107$^{***}$ & -0.123$^{***}$ \\
  & (0.001) & (0.001) & (0.002) & (0.001) \\
 C(Month)[T.9] & -0.171$^{***}$ & -0.130$^{***}$ & -0.094$^{***}$ & -0.104$^{***}$ \\
  & (0.001) & (0.001) & (0.001) & (0.001) \\
 C(Weekday)[T.1] & 0.002$^{***}$ & 0.004$^{***}$ & 0.019$^{***}$ & 0.002$^{***}$ \\
  & (0.000) & (0.001) & (0.001) & (0.001) \\
 C(Weekday)[T.2] & 0.001$^{**}$ & 0.006$^{***}$ & 0.021$^{***}$ & 0.002$^{**}$ \\
  & (0.000) & (0.001) & (0.001) & (0.001) \\
 C(Weekday)[T.3] & 0.003$^{***}$ & 0.004$^{***}$ & 0.021$^{***}$ & 0.004$^{***}$ \\
  & (0.000) & (0.001) & (0.001) & (0.001) \\
 C(Weekday)[T.4] & -0.005$^{***}$ & -0.017$^{***}$ & -0.014$^{***}$ & 0.009$^{***}$ \\
  & (0.000) & (0.001) & (0.001) & (0.001) \\
 C(Weekday)[T.5] & -0.063$^{***}$ & -0.109$^{***}$ & -0.155$^{***}$ & -0.061$^{***}$ \\
  & (0.000) & (0.001) & (0.001) & (0.001) \\
 C(Weekday)[T.6] & 0.001$^{}$ & -0.008$^{***}$ & -0.178$^{***}$ & -0.092$^{***}$ \\
  & (0.001) & (0.001) & (0.001) & (0.001) \\
 CDD1 & 0.003$^{***}$ & 0.000$^{}$ & 0.006$^{***}$ & 0.005$^{***}$ \\
  & (0.000) & (0.001) & (0.001) & (0.001) \\
 CDD2 & -0.001$^{}$ & -0.000$^{}$ & -0.007$^{***}$ & 0.001$^{}$ \\
  & (0.001) & (0.001) & (0.002) & (0.001) \\
 CDD3 & -0.001$^{}$ & 0.001$^{**}$ & 0.004$^{***}$ & -0.003$^{***}$ \\
  & (0.001) & (0.000) & (0.001) & (0.001) \\
 CDD4 & 0.005$^{**}$ & 0.005$^{***}$ & -0.012$^{***}$ & 0.003$^{**}$ \\
  & (0.002) & (0.001) & (0.002) & (0.001) \\
 CDD5 & -0.003$^{**}$ & -0.016$^{***}$ & -0.001$^{}$ & 0.003$^{***}$ \\
  & (0.001) & (0.002) & (0.002) & (0.001) \\
 HDD1 & 0.008$^{***}$ & 0.006$^{***}$ & 0.003$^{***}$ & 0.005$^{***}$ \\
  & (0.000) & (0.000) & (0.000) & (0.000) \\
 HDD2 & 0.005$^{***}$ & 0.005$^{***}$ & 0.002$^{***}$ & -0.000$^{}$ \\
  & (0.000) & (0.000) & (0.000) & (0.000) \\
 HDD3 & 0.000$^{***}$ & 0.003$^{***}$ & -0.000$^{}$ & 0.002$^{***}$ \\
  & (0.000) & (0.000) & (0.000) & (0.000) \\
 HDD4 & 0.002$^{***}$ & -0.000$^{}$ & 0.001$^{*}$ & 0.003$^{***}$ \\
  & (0.000) & (0.000) & (0.000) & (0.000) \\
 HDD5 & 0.001$^{***}$ & 0.002$^{***}$ & 0.002$^{***}$ & -0.000$^{***}$ \\
  & (0.000) & (0.000) & (0.000) & (0.000) \\
 Holiday[T.True] & -0.082$^{***}$ & -0.117$^{***}$ & -0.152$^{***}$ & -0.124$^{***}$ \\
  & (0.001) & (0.001) & (0.001) & (0.001) \\
 Intercept & -59.023$^{***}$ & 43.439$^{***}$ & 25.756$^{***}$ & 153.423$^{***}$ \\
  & (0.617) & (0.580) & (2.611) & (2.715) \\
 lnGDP & 1.253$^{***}$ & 0.302$^{***}$ & 0.518$^{***}$ & 0.541$^{***}$ \\
  & (0.014) & (0.008) & (0.014) & (0.008) \\
 lnPOP & 3.453$^{***}$ & -2.362$^{***}$ & -1.550$^{***}$ & -9.768$^{***}$ \\
  & (0.031) & (0.031) & (0.161) & (0.170) \\
 t & -0.000$^{***}$ & 0.000$^{***}$ & -0.000$^{}$ & 0.000$^{***}$ \\
  & (0.000) & (0.000) & (0.000) & (0.000) \\
\hline \\[-1.8ex]
 Observations & 185,088 & 168,883 & 126,755 & 140,328 \\
 $R^2$ & 0.942 & 0.937 & 0.898 & 0.832 \\
 Adjusted $R^2$ & 0.942 & 0.937 & 0.898 & 0.832 \\
 Residual Std. Error & 0.054(df = 185033) & 0.056(df = 168828) & 0.071(df = 126700) & 0.073(df = 140273)  \\
 F Statistic & 55654.711$^{***}$ (df = 54.0; 185033.0) & 46775.832$^{***}$ (df = 54.0; 168828.0) & 20597.994$^{***}$ (df = 54.0; 126700.0) & 12827.298$^{***}$ (df = 54.0; 140273.0) \\
\hline
\hline \\[-1.8ex]
\textit{Note:} & \multicolumn{4}{r}{$^{*}$p$<$0.1; $^{**}$p$<$0.05; $^{***}$p$<$0.01} \\
\end{tabular}
\end{adjustbox}
\end{table}

\section{Appendix: Country-specific Results}
\label{appendix:b}

\begin{figure}[tbh!]
  \centering
  \includegraphics[width=\textwidth-3em]{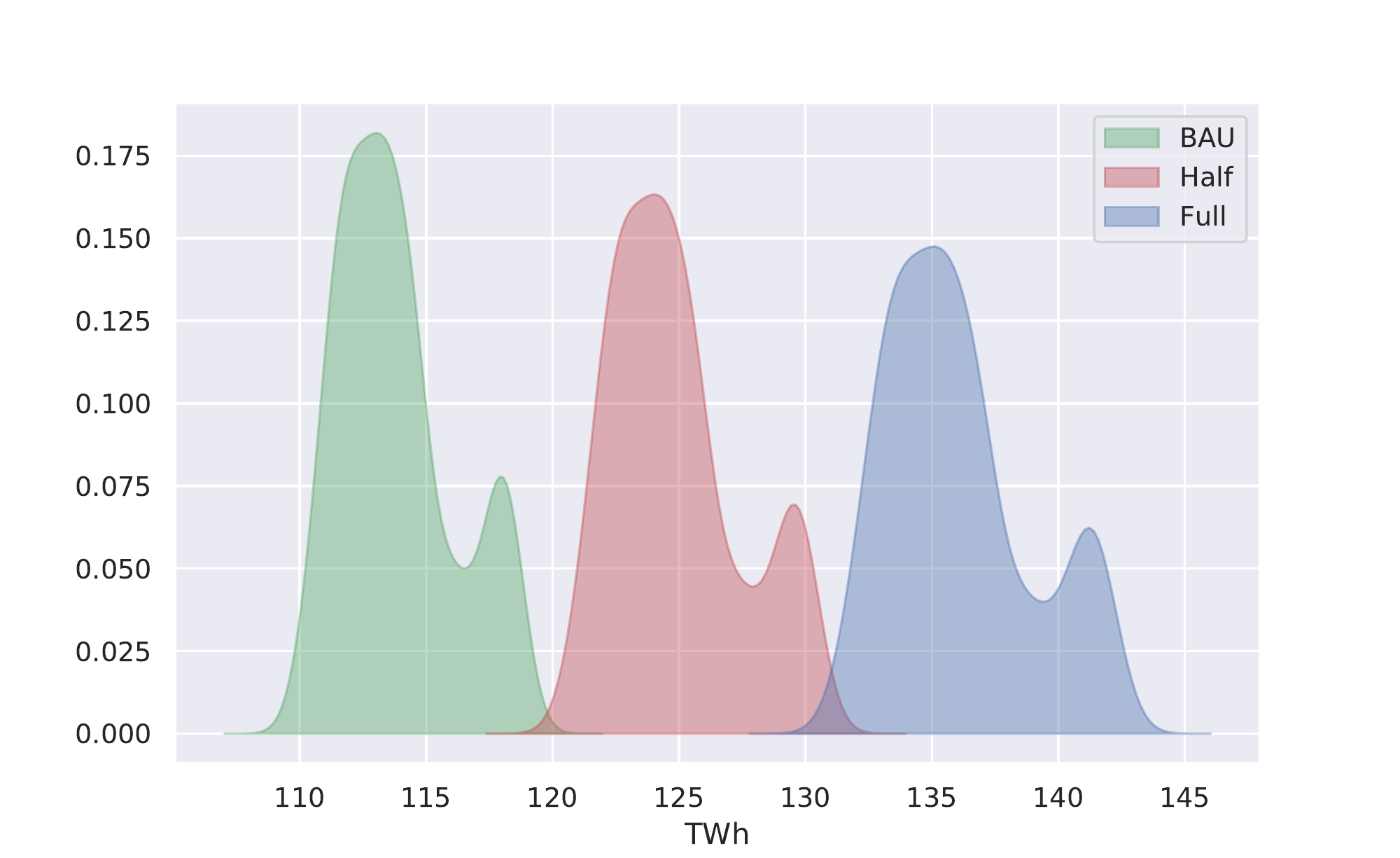}
  \caption{Projected densities of the projected total annual electricity
    consumption in Norway in 2040 for the three heating electrification
    scenarios (business-as-usual, half and full replacement of fossil fuels by
    electricity), under varying weather conditions.}
\end{figure}
\begin{figure}[tbh!]
    \centering
    \includegraphics[width=\textwidth-3em]{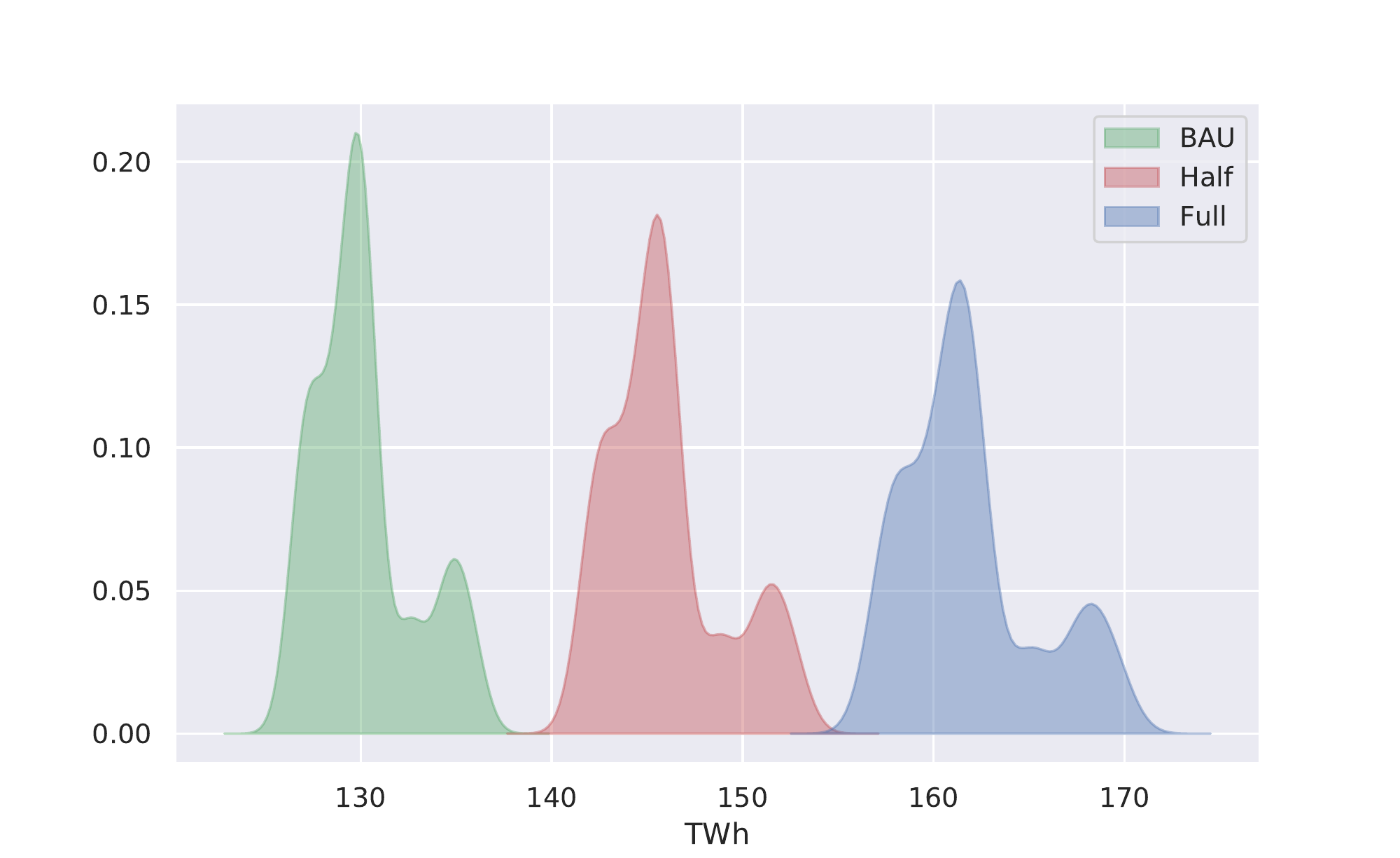}
    \caption{Projected densities of the projected total annual electricity
      consumption in Sweden in 2040 for the three heating electrification
      scenarios (business-as-usual, half and full replacement of fossil fuels by
      electricity), under varying weather conditions.}
\end{figure}
\begin{figure}[tbh!]
  \centering
  \includegraphics[width=\textwidth-3em]{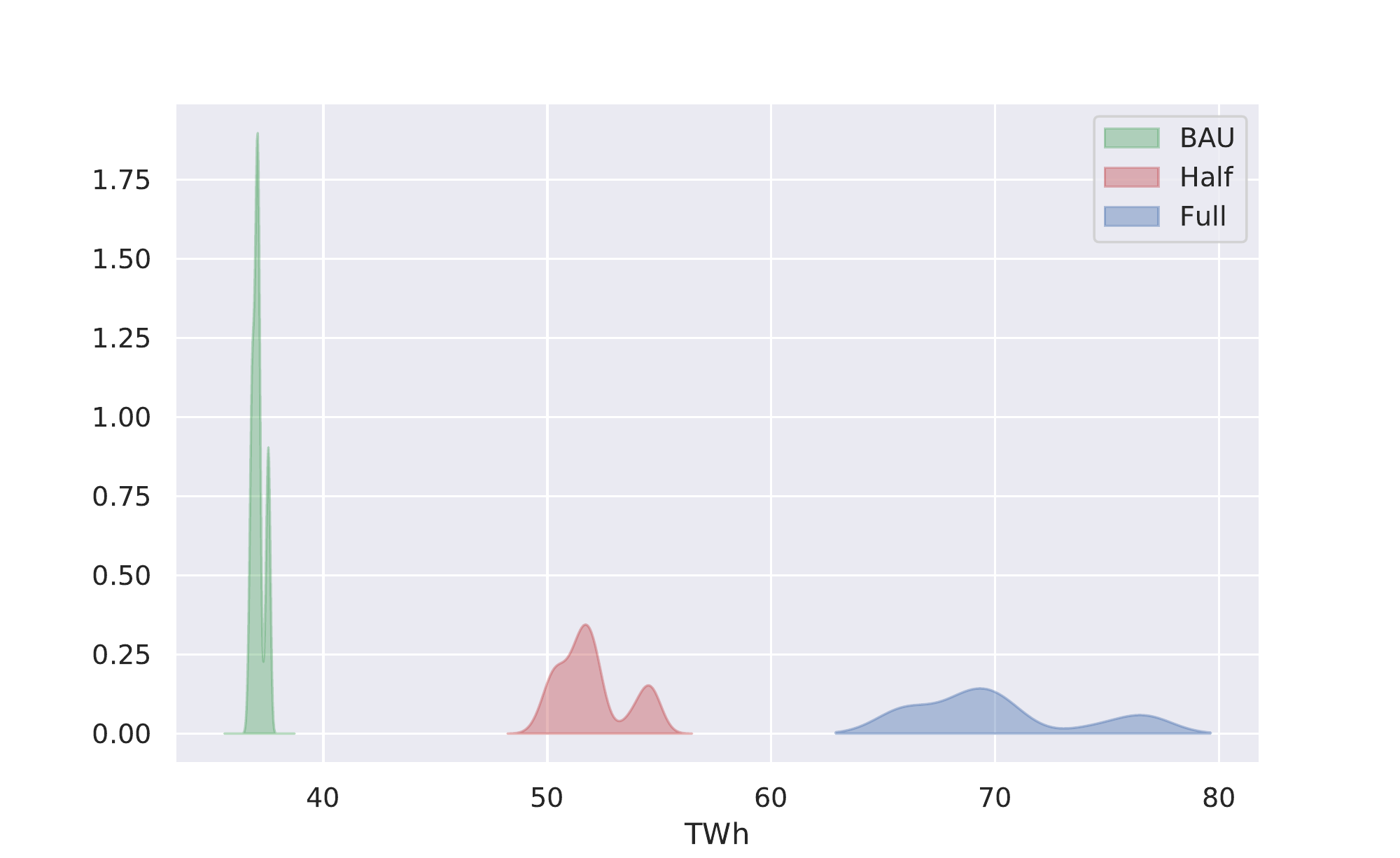}
  \caption{Projected densities of the projected total annual electricity
    consumption in Denmark in 2040 for the three heating electrification
    scenarios (business-as-usual, half and full replacement of fossil fuels by
    electricity), under varying weather conditions.}
\end{figure}
\begin{figure}[tbh!]
  \centering
  \includegraphics[width=\textwidth-3em]{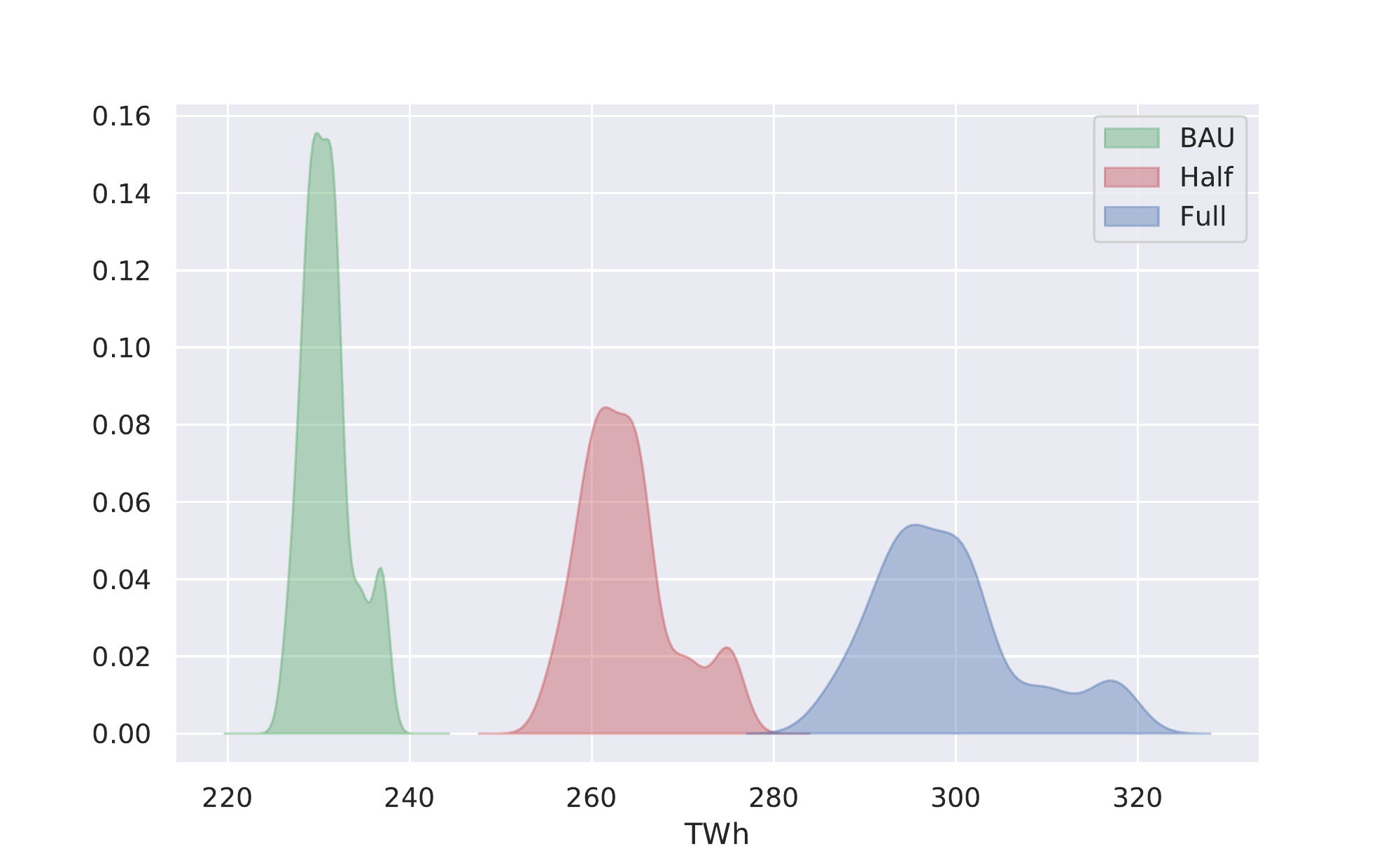}
  \caption{Projected densities of the projected total annual electricity
    consumption in Finland in 2040 for the three heating electrification
    scenarios (business-as-usual, half and full replacement of fossil fuels by
    electricity), under varying weather conditions.}
\end{figure}

\begin{table}[tbh!]
  \centering
  \caption{Summary of the projected total consumption for 2040 for the three
    heating electrification scenarios.}
  \label{tab:total}
  \begin{tabular}{|l|r|r|r|}
    \hline
    \multicolumn{4}{|c|}{\textbf{Business-As-Usual Scenario}} \\
    \hline
    \textbf{Country} & \textbf{Mean} (TWh) & \textbf{Std. dev.} (TWh) &
    \textbf{CVaR}$_{95\%}$ (TWh) \\
    \hline
    Norway & 113.9 & 2.3 & 118.4 \\
    Sweden & 130.1 & 2.6 & 135.7 \\
    Denmark & 37.1 & 0.3 & 37.6 \\
    Finland & 231.0 & 2.7 & 237.2 \\
    \textbf{Nordic} & 512.1 & 7.7 & 528.9 \\
    \hline
    \multicolumn{4}{|c|}{\textbf{Half Electrification Scenario}} \\
    \hline
    \textbf{Country} & \textbf{Mean} (TWh) & \textbf{Std. dev.} (TWh) &
    \textbf{CVaR}$_{95\%}$ (TWh) \\
    \hline
    Norway & 124.9 & 2.5 & 130.0 \\
    Sweden & 146.0 & 3.0 & 152.5 \\
    Denmark & 52.0 & 1.5 & 54.8 \\
    Finland & 263.9 & 5.1 & 275.6 \\
    \textbf{Nordic} & 586.7 & 11.9 & 612.9 \\
    \hline
    \multicolumn{4}{|c|}{\textbf{Full Electrification Scenario}} \\
    \hline
    \textbf{Country} & \textbf{Mean} (TWh) & \textbf{Std. dev.} (TWh) &
    \textbf{CVaR}$_{95\%}$ (TWh) \\
    \hline
    Norway & 136.1 & 2.8 & 141.8 \\
    Sweden & 161.9 & 3.4 & 169.4 \\
    Denmark & 70.1 & 3.7 & 77.2 \\
    Finland & 299.2 & 8.1 & 318.3 \\
    \textbf{Nordic} & 667.2 & 17.7 & 706.6 \\
    \hline
  \end{tabular}
\end{table}

\begin{figure}[tbh!]
  \centering
  \includegraphics[width=\textwidth-3em]{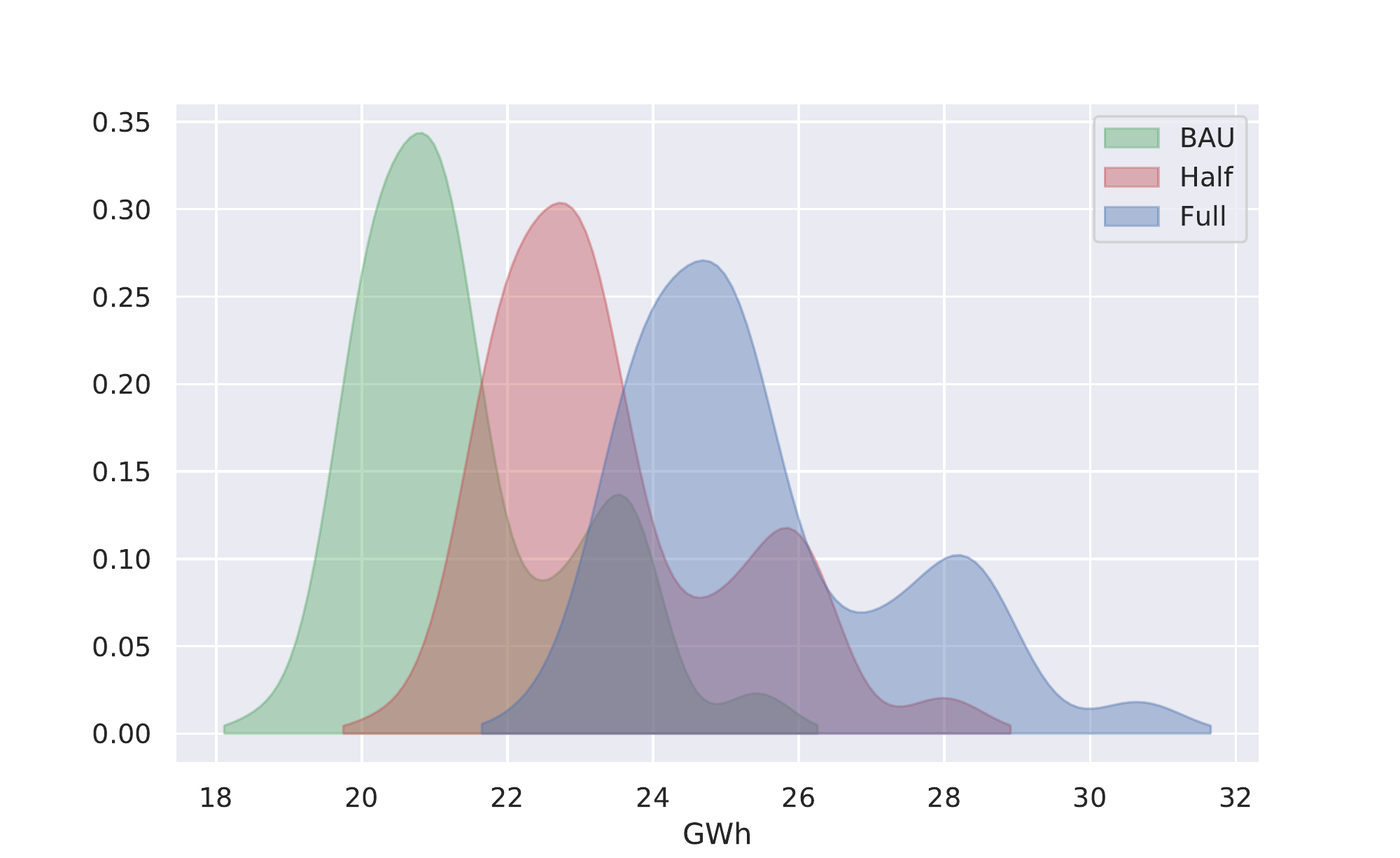}
  \caption{Projected densities of the Norwegian electricity consumption in the
    peak hour in 2040 for the three heating electrification scenarios
    (business-as-usual, half and full replacement of fossil fuels by
    electricity), under varying weather conditions.}
\end{figure}  
\begin{figure}[tbh!]
    \centering
    \includegraphics[width=\textwidth-3em]{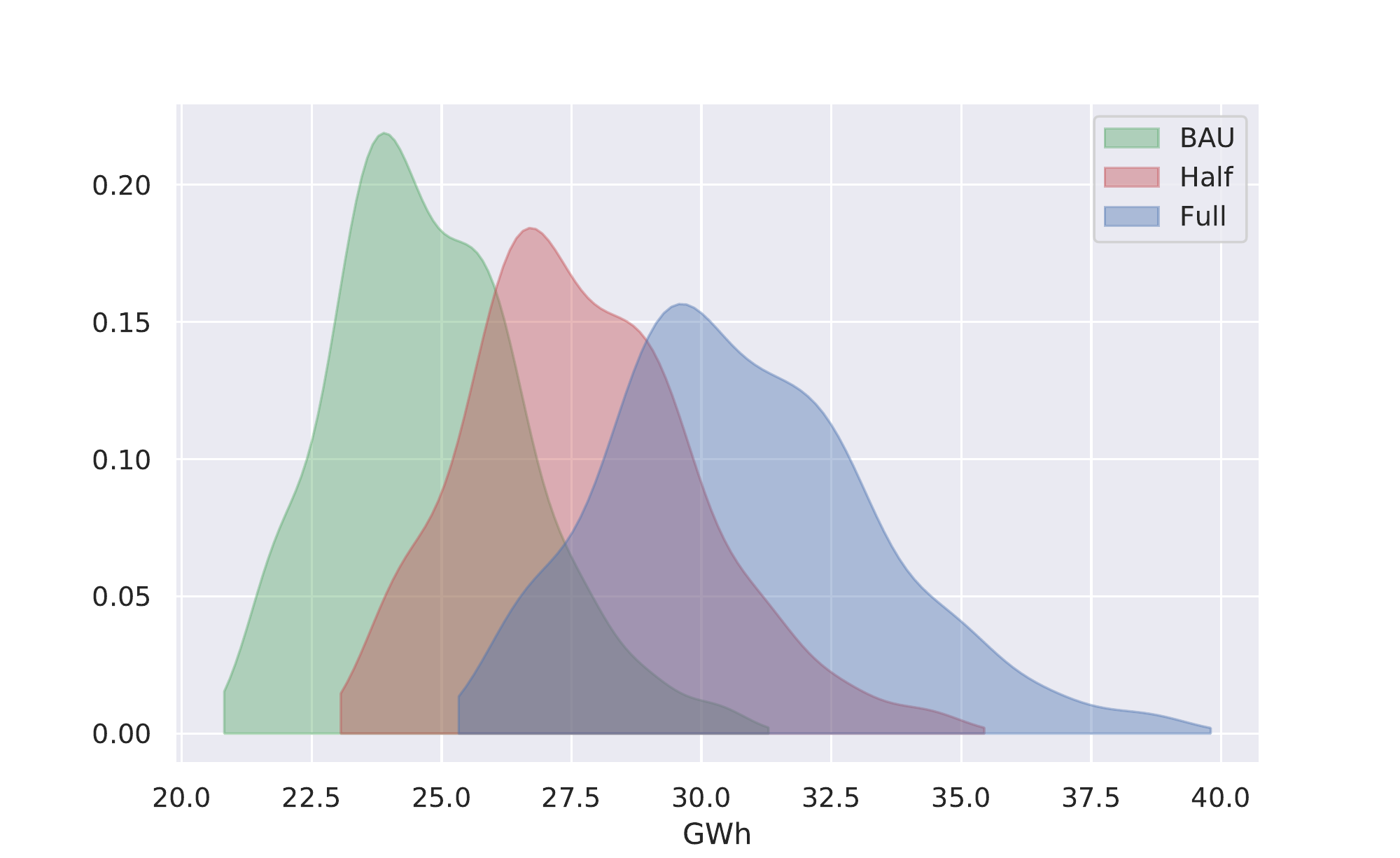}
    \caption{Projected densities of the Swedish electricity consumption in the
      peak hour in 2040 for the three heating electrification scenarios
      (business-as-usual, half and full replacement of fossil fuels by
      electricity), under varying weather conditions.}
\end{figure}
\begin{figure}[tbh!]
  \centering
  \includegraphics[width=\textwidth-3em]{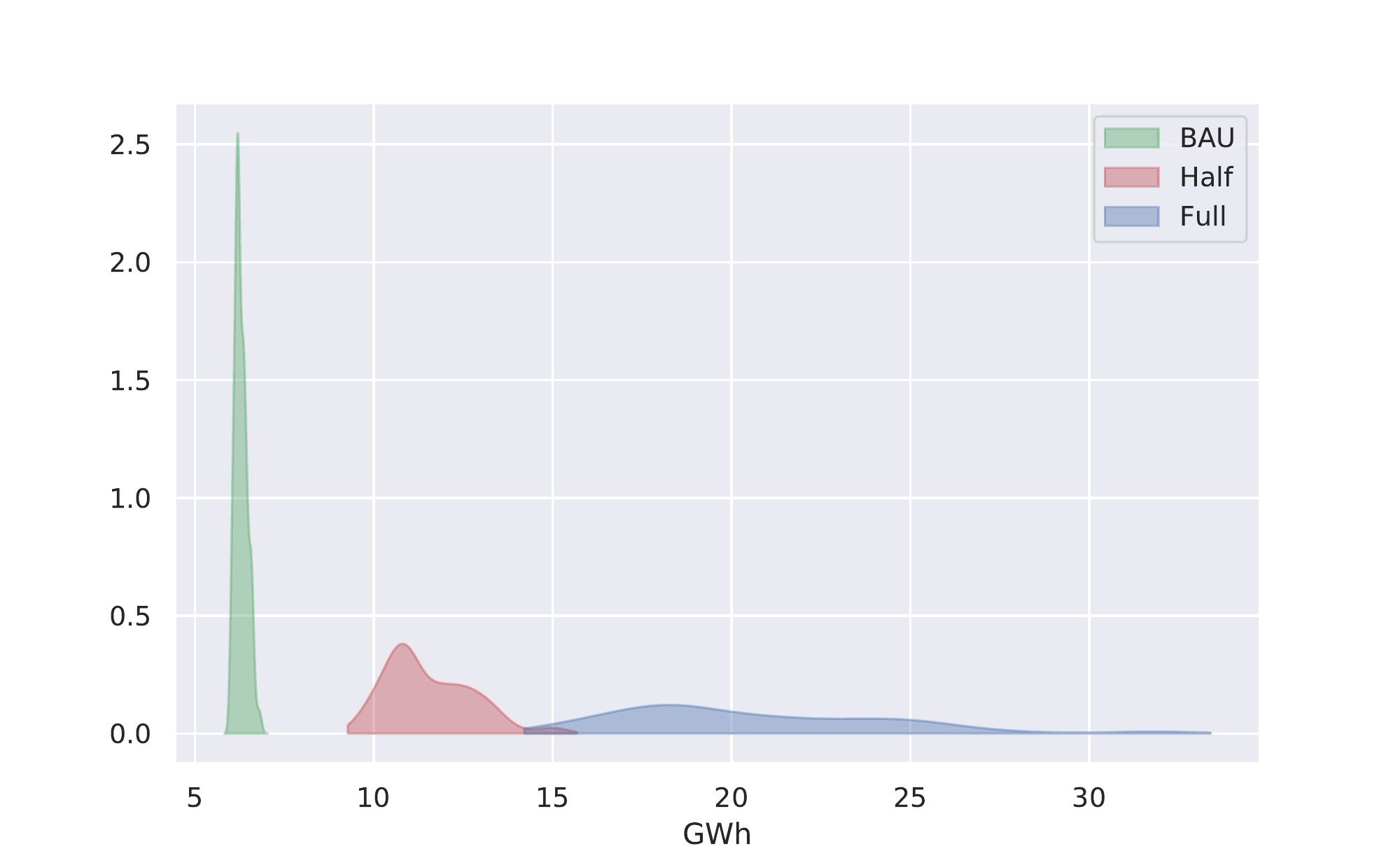}
  \caption{Projected densities of the Danish electricity consumption in the peak
    hour in 2040 for the three heating electrification scenarios
    (business-as-usual, half and full replacement of fossil fuels by
    electricity), under varying weather conditions.}
\end{figure}
\begin{figure}[tbh!]
  \centering
  \includegraphics[width=\textwidth-3em]{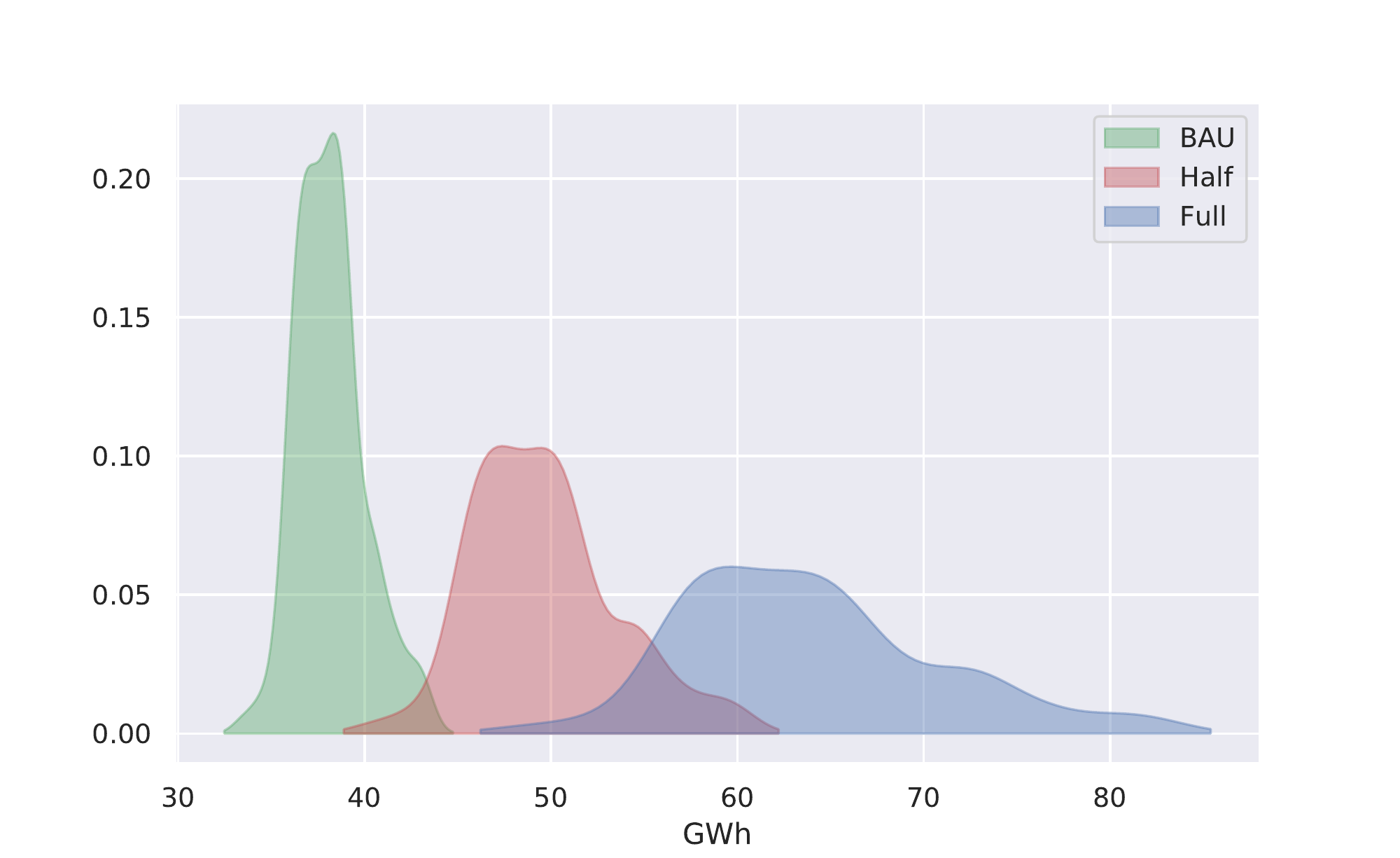}
  \caption{Projected densities of the Finnish electricity consumption in the
    peak hour in 2040 for the three heating electrification scenarios
    (business-as-usual, half and full replacement of fossil fuels by
    electricity), under varying weather conditions.}
\end{figure}

\begin{table}[tbh!]
  \centering
  \caption{Summary of the projected consumption during the peak hour in 2040 for
    the three heating electrification scenarios.}
  \label{tab:peakcons}
  \begin{tabular}{|l|r|r|r|}
    \hline
    \multicolumn{4}{|c|}{\textbf{Business-As-Usual Scenario}} \\
    \hline
    \textbf{Country} & \textbf{Mean} (GWh) & \textbf{Std. dev.} (GWh) &
    \textbf{CVaR}$_{95\%}$ (GWh) \\
    \hline
    Norway & 21.4 & 1.4 & 24.8 \\
    Sweden & 24.8 & 1.8 & 29.3 \\
    Denmark & 6.3 & 0.2 & 6.7 \\
    Finland & 38.2 & 1.8 & 42.9 \\
    \textbf{Nordic} & 88.6 & 5.0 & 102.8 \\
    \hline
    \multicolumn{4}{|c|}{\textbf{Half Electrification Scenario}} \\
    \hline
    \textbf{Country} & \textbf{Mean} (GWh) & \textbf{Std. dev.} (GWh) &
    \textbf{CVaR}$_{95\%}$ (GWh) \\
    \hline
    Norway & 23.4 & 1.6 & 27.2 \\
    Sweden & 27.7 & 2.2 & 33.1 \\
    Denmark & 11.5 & 1.2 & 14.5 \\
    Finland & 49.6 & 3.8 & 59.4 \\
    \textbf{Nordic} & 108.8 & 8.4 & 133.0 \\
    \hline
    \multicolumn{4}{|c|}{\textbf{Full Electrification Scenario}} \\
    \hline
    \textbf{Country} & \textbf{Mean} (GWh) & \textbf{Std. dev.} (GWh) &
    \textbf{CVaR}$_{95\%}$ (GWh) \\
    \hline
    Norway & 25.5 & 1.8 & 29.8 \\
    Sweden & 30.8 & 2.6 & 37.1 \\
    Denmark & 20.5 & 3.7 & 30.3 \\
    Finland & 63.9 & 6.8 & 81.3 \\
    \textbf{Nordic} & 134.8 & 14.3 & 176.6 \\
    \hline
  \end{tabular}
\end{table}

\begin{figure}[tbh!]
  \centering
  \includegraphics[width=\textwidth]{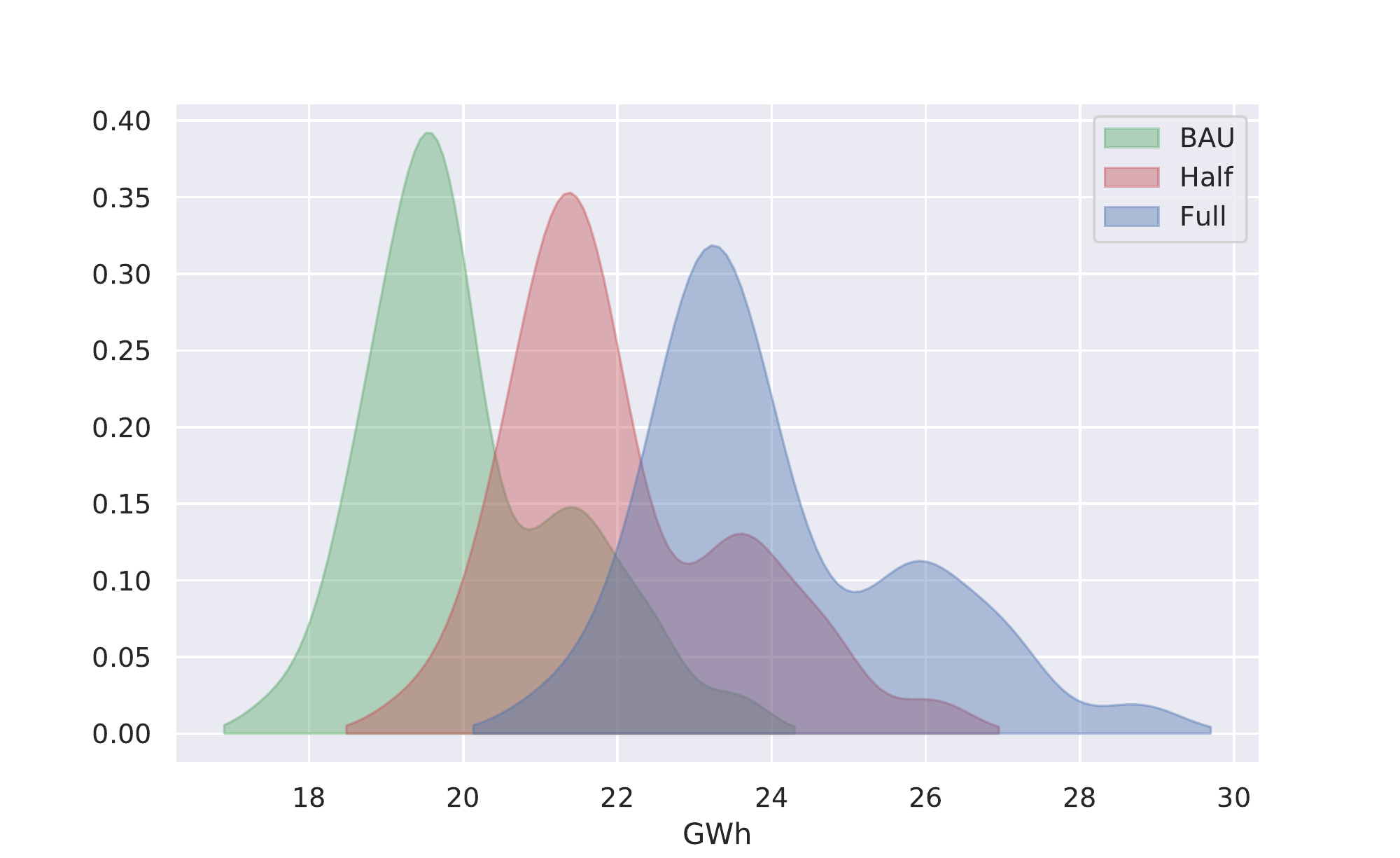}
  \caption{Projected densities of the Norwegian residual electricity demand in
    the peak hour in 2040 for the three electrification scenarios
    (business-as-usual, half and full replacement of fossil fuels by
    electricity), under varying weather conditions.}
\end{figure}  
\begin{figure}[tbh!]
    \centering
    \includegraphics[width=\textwidth]{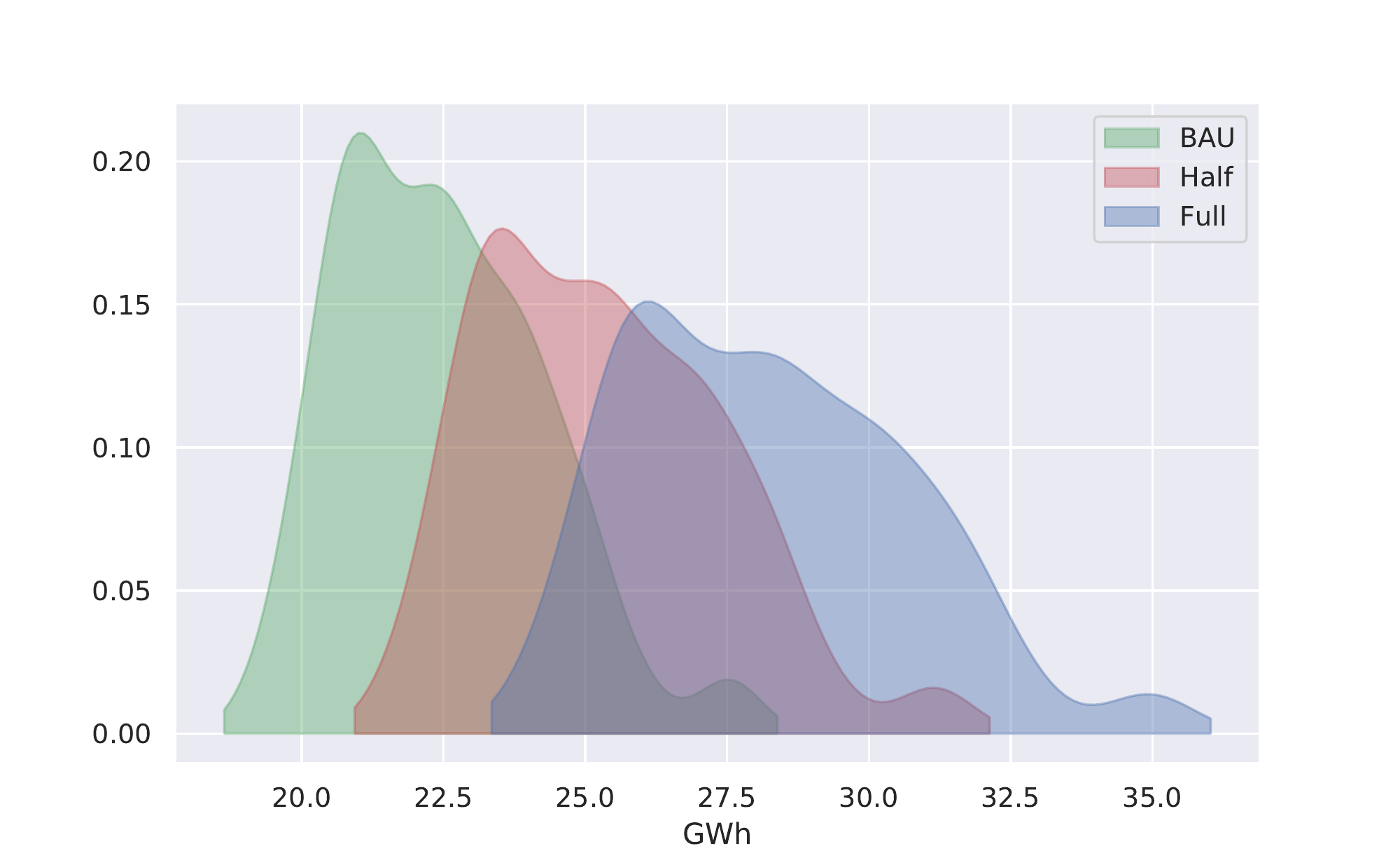}
    \caption{Projected densities of the Swedish residual electricity demand in
      the peak hour in 2040 for the three electrification scenarios
      (business-as-usual, half and full replacement of fossil fuels by
      electricity), under varying weather conditions.}
\end{figure}
\begin{figure}[tbh!]
  \centering
  \includegraphics[width=\textwidth]{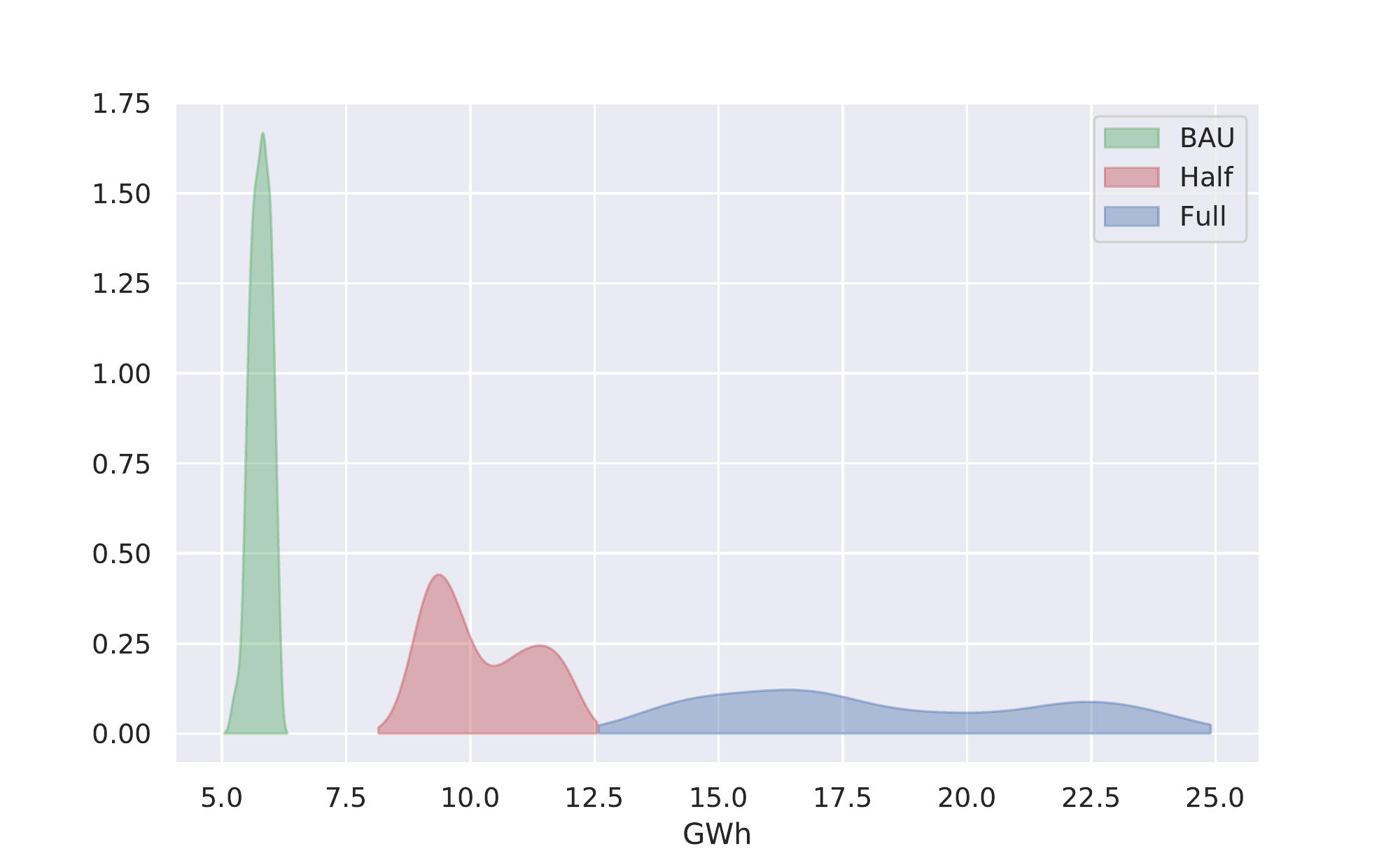}
  \caption{Projected densities of the Danish residual electricity demand in the
    peak hour in 2040 for the three electrification scenarios
    (business-as-usual, half and full replacement of fossil fuels by
    electricity), under varying weather conditions.}
\end{figure}
\begin{figure}[tbh!]
  \centering
  \includegraphics[width=\textwidth]{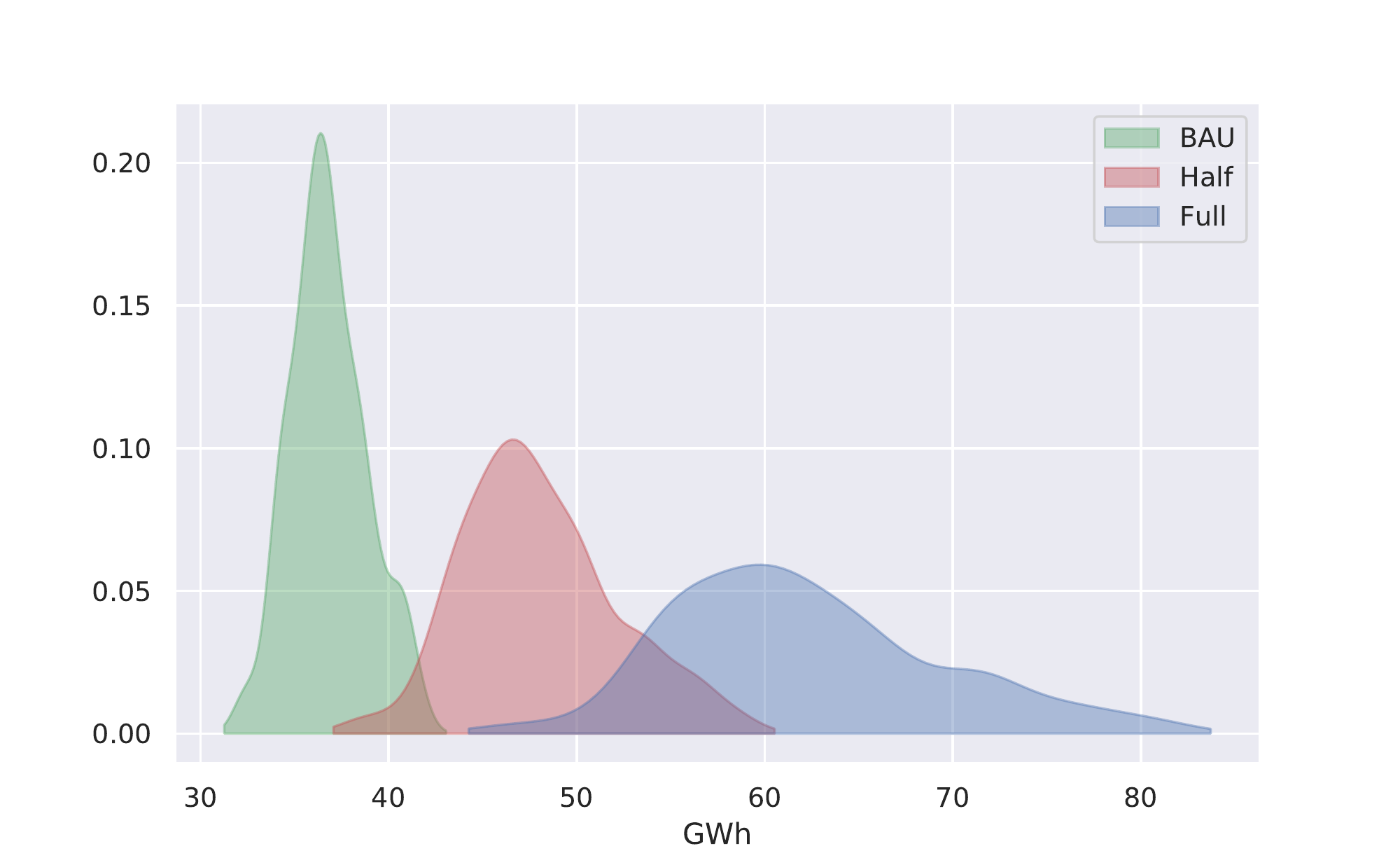}
  \caption{Projected densities of the Finnish residual electricity demand in the
    peak hour in 2040 for the three electrification scenarios
    (business-as-usual, half and full replacement of fossil fuels by
    electricity), under varying weather conditions.}
\end{figure}

\begin{table}[tbh!]
  \centering
  \caption{Summary of the projected residual demand during the peak hour in 2040
    for the three heating electrification scenarios.}
  \label{tab:peakres}
  \begin{tabular}{|l|r|r|r|}
    \hline
    \multicolumn{4}{|c|}{\textbf{Business-As-Usual Scenario}} \\
    \hline
    \textbf{Country} & \textbf{Mean} (GWh) & \textbf{Std. dev.} (GWh) &
    \textbf{CVaR}$_{95\%}$ (GWh) \\
    \hline
    Norway & 20.1 & 1.3 & 23.0 \\
    Sweden & 22.5 & 1.8 & 26.7 \\
    Denmark & 5.8 & 0.2 & 6.1 \\
    Finland & 36.8 & 2.0 & 41.1 \\
    \textbf{Nordic} & 79.7 & 5.0 & 92.5 \\
    \hline
    \multicolumn{4}{|c|}{\textbf{Half Electrification Scenario}} \\
    \hline
    \textbf{Country} & \textbf{Mean} (GWh) & \textbf{Std. dev.} (GWh) &
    \textbf{CVaR}$_{95\%}$ (GWh) \\
    \hline
    Norway & 22.0 & 1.5 & 25.5 \\
    Sweden & 25.3 & 2.1 & 30.2 \\
    Denmark & 10.2 & 1.0 & 12.1 \\
    Finland & 48.0 & 4.1 & 57.4 \\
    \textbf{Nordic} & 98.9 & 8.1 & 118.4 \\
    \hline
    \multicolumn{4}{|c|}{\textbf{Full Electrification Scenario}} \\
    \hline
    \textbf{Country} & \textbf{Mean} (GWh) & \textbf{Std. dev.} (GWh) &
    \textbf{CVaR}$_{95\%}$ (GWh) \\
    \hline
    Norway & 24.1 & 1.7 & 28.0 \\
    Sweden & 28.3 & 2.4 & 33.9 \\
    Denmark & 18.4 & 3.3 & 24.0 \\
    Finland & 62.2 & 7.1 & 79.3 \\
    \textbf{Nordic} & 123.8 & 13.4 & 154.8 \\
    \hline
  \end{tabular}
\end{table}

\end{document}